\documentclass[12pt]{iopart}

\expandafter\let\csname equation*\endcsname\relax
\expandafter\let\csname endequation*\endcsname\relax
\usepackage{amsmath}
\usepackage{graphics,epsfig,placeins,subfig,wrapfig}
\usepackage{subfig}
\usepackage{color,soul}

\begin{document}

\title[First steps on modelling wave propagation in isotropic-heterogeneous media]{First steps on modelling wave propagation in isotropic-heterogeneous media: Numerical simulation of P-SV waves}

\author{Gabriela Landinez, 
Santiago Rueda,
Fabio D. Lora-Clavijo}
\address{Escuela de F\'isica, Universidad Industrial de Santander, A. A. 678,
Bucaramanga 680002, Colombia.}
\ead{fadulora@uis.edu.co}

\vspace{10pt}
\begin{indented}
\item[]May 2021
\end{indented}

\begin{abstract}
In geophysics, wave propagation in elastic media is a crucial subject. In this context, seismology has made significant progress as a result of numerous advances, among these stands out the advancement of numerical methods such as the finite-difference one. Usually, seismic wave propagation in elastic media results in complex systems of partial differential equations, whose solutions cannot be obtained in an analytical way, especially when dealing with heterogeneous media. In consequence, there exists a necessity to implement numerical methods. However, the available information about the construction of a numerical solution of these equations is not as explicit as it should be. Our main goal is to provide pedagogical instructions for undergraduate students who want to improve their computational skills as well as their knowledge of the subject. The last through current teaching methods involving challenging problems and transversal tools. Particularly, we provide a brief description of the equations and a detailed description of the numerical solution of the seismic wave equation. Furthermore, we model two different seismic explosive sources in both homogeneous and heterogeneous media as an illustrative example. In the results, we include velocity maps showing wave propagation in the x-z plane, and the z-velocity as a function of time measured in a series of detectors distributed in the numerical domain. 
\end{abstract}

\vspace{2pc}

\section{Introduction}

The development of skills, to solve challenges linking  computational tools, is nowadays a necessary complement to analytical studies of physical phenomena expressed through laws based on experimental physics. Many authors agree with the idea that to properly acquire computational skills, students need more than programming classes, they need to face challenging problems that allows not only to learn the computational concepts, but also the fundamental principles of physics and how to apply them \cite{Yasar2014, Landau2006}. Teaching nowadays should not only focus on theoretical learning, but also in the inclusion of transversal tools, like numerical methods, data analysis, among others, that allows the student to face more realist problems. When students face these situations they have to move back an forth between supplying some evidence, and given more premises that allows to reach a logical conclusion. 

Nowadays, the numerical approach enables solving complex problems in physics related to systems of equations that cannot be solved analytically. In particular, wave propagation in elastic media is modelled with systems of equations that are not easily solved when dealing with heterogeneous media, that is, stratified media, homogeneous in each of its layers. In this case, it is necessary taking into account changes in the speed of each type of waves, and of course, reflection and transmission at the interface between layers. This topic is fundamental for seismic wave propagation and is associated with the concepts of stress and strain usually discussed in introductory physics courses. Seismology is the science that studies these waves, their sources, their propagation, and what they tell about the structure of Earth and the physics of earthquakes \cite{Shearer2019}. The study of seismic waves requires analysis of wave propagation in complex media, such as the interface between two different materials and the interpretation of data contained in seismographs. Moreover, concepts developed for this study are nowadays applied to different areas of knowledge, for instance, helioseismology, which is responsible for studying the structure and dynamics of the Sun through its oscillations \cite{gough1996perspectives, tayler1997sun, Aki2002}.

Although seismology is a very young branch of science, thanks to technological advances, which allow the improvement of seismographs and other data acquisition systems, considerable advances have been made in this area. On the other hand, data accessibility requires the improvement in data processing techniques as well as the development of numerical methods. Numerical modelling has had significant growth in recent years and is the key in many physical research fields. Several seismic modelling approaches have been developed, with different complexity levels of numerical schemes. The most popular are: finite-difference method, finite-element method, pseudo-spectral methods, among others \cite{virieux2011review}. In particular, the finite-difference method has shown to be efficient enough to solve problems of elastic mechanics and to fully describe wave propagation in media with spatial variations of elastic properties \cite{Feng2013, favorskaya2018modelling}. 

The finite-difference method was one of the first approaches applied to the numerical solution of partial differential equations (PDEs). The mathematician Runge, C. \cite{Runge1908} was probably the first one to apply this method by solving torsional problems, reducing them to a system of linear equations \cite{Timoshenko1970}. Few years later an advance was present by Richardson, L. \cite{Richardson1910} who employed this method for the solution of Laplace equations and some equations related to oscillations of a given media. Through an iterative process he obtained approximate values of the stresses on dams due to gravity and water pressure, also emphasizing that the method is not limited to this equations but can be used in many others, like the elastodynamic equations \cite{Moczo2007, Robertsson2020}. The principle of the finite-difference method is to employ a Taylor series expansion to discretize the derivative operators involved in the PDE. This approximates derivatives by combining neighbouring function values on the grid, where, as it was mentioned before, the combination is derived using the Taylor expansion of the function at the different grid points. By applying this method, the continuous domain is discretized and the differential terms of the equation are converted into a linear algebraic expressions. 

Seismo-elastic waves that propagate through an isotropic medium consist of two components: the P-waves and SV-waves, commonly called the P-SV wave system \cite{muller2007theory}. P-waves are pressure or compressional waves and travel as elastic motions at the highest speeds. They are longitudinal waves that can be transmitted by both solid and liquid materials in the Earth’s interior. On the contrary, SV-waves are shear waves and travels at the slowest speeds. The particles in SV-waves move perpendicular to the direction in which the wave propagates and can only travel in solid materials \cite{muller2007theory}. 

The study of P-SV waves using finite-difference methods has a long trajectory, which lately has grown thanks to the computational improvements. Currently, it is among the most commonly used methods for simulating wave propagation in complex media \cite{Moczo2007, Robertsson2020}, but other techniques such as spectral methods are also commonly used. For further information, De Basabe and Sen \cite{deBasabe2015} provide a comparison between the finite-difference and spectral element methods for elastic wave propagation in media with a fluid-solid interface. According to these authors, two different approaches have been used to perform simulations in heterogeneous media.  First, the partitioned approach, in which a different system of equations is used in each phase with internal interface boundary conditions. Second, the monolithic approach, in which the PDEs are the same and there are no explicit interface boundary conditions because they are implicitly imposed by the change in media parameters. The absence of internal boundary conditions makes easier the numerical implementation and that is the reason why the monolithic approach is one of the most common in geophysics \cite{deBasabe2015}. Moreover, the preference for this approach is justified by the use of the staggered-grid finite-difference method. Virieux \cite{Virieux1986} showed that the elastodynamic equations can be solved by this finite-difference technique, which uses velocity and stress as physical quantities distributed on a staggered grid. In particular, he employed a Gaussian pulse as an explosive source by adding a value to the stress terms. Other sources, like the derivative of a Gaussian pulse and Ricker wavelets of different frequencies, have been implemented also by Virieux \cite{Virieux1986} and other authors \cite{Vossen2002, Hong2003, Serdyukov2019}. Concerning to the implementation of external boundary conditions, one problem everyone has to face is associated with the reflection of waves within the numerical domain. A possible solution for this, proposed by Lindman \cite{Lindman1975}, consists of imposing boundary conditions that act as an infinite region of free space, avoiding the annoying reflected waves. Cao \cite{Cao1992}, working with the  Lindman free space, together with a dissipation zone to absorbing boundaries, optimized the result.

Even with all the advances made in the field of seismology, the problem of seismic wave propagation is not completely solved. For example, despite that near-surface media is considered the easiest media to image due to its proximity to seismic sensors, it is actually one of the most difficult to study because near-surface media have abrupt changes in the seismic properties \cite{Lay2009}. Also, this media have very high attenuation, which causes amplitude decay and phase distortion of the seismic waves. When attenuation varies with direction another challenge arise: attenuation anisotropy. Seismic attenuation anisotropy has a significant impact on the propagation of seismic waves, therefore, seismic wave simulation in the presence of anisotropic attenuation is needed \cite{Qiao2020}. Although anisotropic models are currently used in worldwide seismology, parameter estimation for complex anisotropic models remains a big challenge for experts \cite{Ivanov2019}. 

As we have seen, many challenges arise every day, and with them,  new improvements in numerical simulation techniques and a better performance of computer systems are searched to affront them. Recently the use of supercomputers to simulate large-scale 3D geophysical medium has increased intensively \cite{Karavaev2015, Khokhlov2016}, as a key to handle with huge computations with large amounts of 3D data. The above allows us to deal with fundamental problems as the studying of the structure of Earth’s crust and upper mantle and geodynamic processes \cite{Kovalevsky2016}, or how basin structures can amplify the destruction effects during an earthquake \cite{Furumura2000, Ba2021}. However, these studies are not only limited at the earth, nowadays simulations of wave propagation in mars has been made \cite{Bozdaug2017}, realizing that all these numerical tools are able to affront new challenges, that include different seismic models and new scenarios. One of them consists in the development of new models of classic seismology combined with glaciology, oceanography, and other fields to understand the causes and consequences of glacial earthquakes \cite{Lay2009}.

From our experience, we have found that the existent literature about the construction of the numerical solution, i.e., the finite-difference method, the initial conditions, boundary conditions and the seismic source, is not as explicit as it may be expected by students having their first contact with this subject. In this educative work, we address the first steps to study wave propagation in heterogeneous media from the numerical approach, which represents a significant advance in the way of teaching physics via challenging problems. There are many academic works in which wave propagation plays an important role, however, most of them focus on just theoretical or experimental means \cite{Wittmann2002, Maulidah2018, Goodhew2019}, leaving aside situations where transversal skills are needed. In this work we go a step forward, not only by including more realistic properties of the media, but also in the modern way to teach a subject in physics using transversal tools.
This is the reason why we present a paper that is very detailed in the construction and implementation of the numerical solution of the wave equation in a complex media. We focus on the numerical solution of the problem and omit some of the mathematical background that is actually very well described in the literature. The paper is organized as follows. In Sec. \ref{sec:equations} we present a brief description of the basic equations for two dimensional P-SV wave propagation. We use the velocity-stress formulation \cite{Virieux1986, Vossen2002} that allows writing the equations for seismic wave propagation as a first-order hyperbolic partial differential system of equations for the stress and particle velocity. In Sec. \ref{sec:NM} and Sec. \ref{sec:IDBC} we present in detail the numerical methods and a description of the code as well as the initial data and the boundary conditions. In Sec. \ref{sec:results} we show the results of the computational simulations based on a finite-difference scheme for seismic waves propagating in a homogeneous isotropic media and through the interface between two different media (heterogeneous-isotropic media). The interface is not treated by any boundary condition, it is simply represented only by changes of density and other elastic parameters. We solve these equations numerically for two different seismic sources: the Ricker and sine wavelets. Finally in Sec. \ref{sec:con} we provide a discussion of the results.

\section{Equations}\label{sec:equations}

Elastic wave dynamics is governed by the equation of motion and the elastic constitutive equation, which relates the stress and strain tensors \cite{Vossen2002}. The basic equations for wave propagation can be formulated in different but equivalent ways \cite{Robertsson2020}. For instance, one can either use the equation of motion, which is a second-order hyperbolic system, or a first-order hyperbolic system for the stress and velocity variables. We will focus our attention on the last way, known as the velocity-stress formulation, which according to Robertsson \cite{Robertsson2020} is by far the most popular for modelling wave propagation in elastic media.

The derivation of the velocity-stress formulation follows easily using the equation of motion without body force terms. These equations provide information on the variation of the stress exerted on a given medium with respect to the directions in which the wave propagates. Moreover, the set of equations allow us to analyze the speed variation of the waves that propagate in different directions. Even if the equations are valid only for homogeneous media, it is possible to model wave propagation for heterogeneous media as long as it is stratified, that is, the media is composed of different phases or layers.  Such modelling is relevant to study the seismic events, which take place in zones where the composition of the medium is mainly dependent on the depth, as the subsoil layers.

\subsection{Strain tensor}

\begin{figure}
\begin{center}
\includegraphics[scale=1.0]{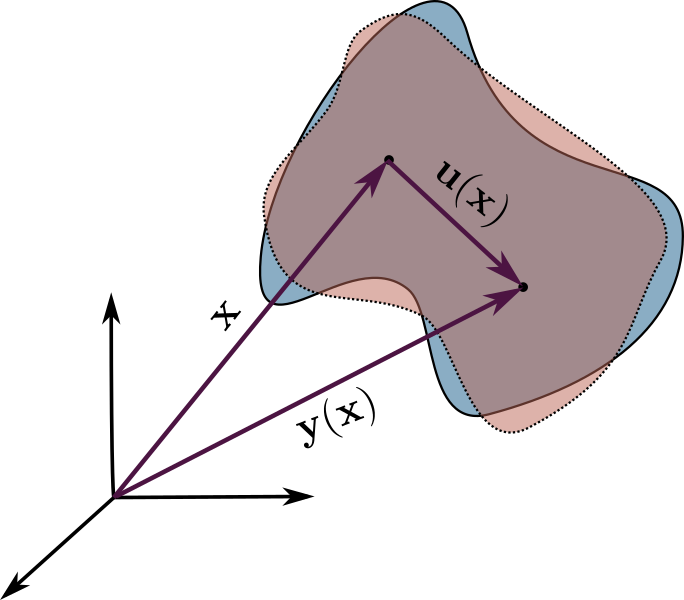}
\caption{Arbitrary volume before (blue shadow) and after (pink shadow) a small deformation. The vector $\mathbf{x}$ denotes the position of an arbitrary point in the volume before the deformation. When the volume is deformed the point suffers a displacement $\mathbf{u}(\mathbf{x})$, and its new position is represented as $\mathbf{y}(\mathbf{x})$.}
\label{Scheme1}
\end{center}
\end{figure}

Consider a point in the solid seen in Figure \ref{Scheme1}, whose position is represented by the vector $\mathbf{x}$. When the object suffers a small deformation, the starting point is shifted to a new position given by $\mathbf{y}(\mathbf{x})$, in such a way that the displacement between these points is $\mathbf{u}(\mathbf{x})=\mathbf{y}(\mathbf{x})-\mathbf{x}$. Considering a cartesian basis of vectors $\{\mathbf{e}_1,\mathbf{e}_2,\mathbf{e}_3\}$, the vectors $\mathbf{x}$, $\mathbf{u}(\mathbf{x})$ and $\mathbf{y}(\mathbf{x})$ can be written as follows
\begin{align}
    \mathbf{x} &= \sum^3_{i=1} x_i \mathbf{e}_i,\\
    \mathbf{u}(\mathbf{x}) &= \sum^3_{i=1} u_i \mathbf{e}_i,\\ 
    \mathbf{y}(\mathbf{x}) &= \sum^3_{i=1} (x_i + u_i) \mathbf{e}_i.
\end{align}

\begin{figure}
\begin{center}
\includegraphics[scale=1.0]{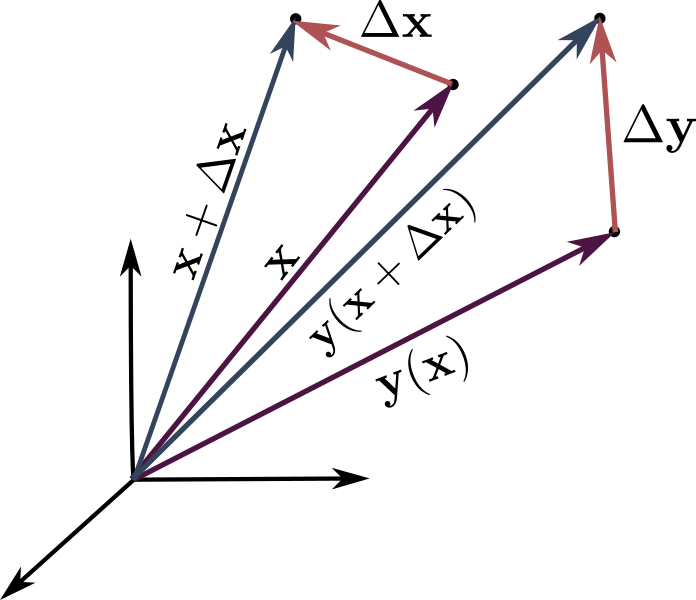} 
\includegraphics[scale=1.0]{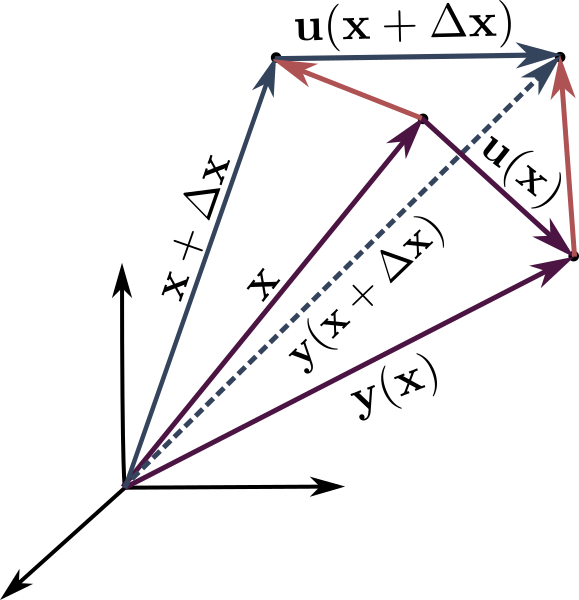} 
\caption{(Left) Nearby points $\mathbf{x}$, $\mathbf{x} + \Delta\mathbf{x}$ (before the deformation) and $\mathbf{y}(\mathbf{x})$, $\mathbf{y}(\mathbf{x} + \Delta\mathbf{x})$ (after the deformation). (Right) Once the volume is deformed, the points $\mathbf{x}$ and $\mathbf{x} + \Delta\mathbf{x}$ suffer a displacement $\mathbf{u}(\mathbf{x})$ and  $\mathbf{u}(\mathbf{x}+\Delta\mathbf{x})$, respectively.}
\label{Scheme2}
\end{center}
\end{figure}

Now, think about an additional point, close to the original one,  at $\mathbf{x} + \Delta \mathbf{x}$. After applying the deformation, this new point is now located at  $\mathbf{y}(\mathbf{x} + \Delta \mathbf{x})$, so that the displacement between these points is given by $\mathbf{u}(\mathbf{x} + \Delta \mathbf{x})$, see Figure \ref{Scheme2}. 
As it can be seen, the difference between both displacements is given by
\begin{align}\label{2.0.0}
    \Delta \mathbf{u}(\mathbf{x}) = \sum^3_{i=1} \Delta u_i \mathbf{e}_i,
\end{align}
being $\Delta u_i$ the components of $\Delta \mathbf{u}$ respect to the cartesian base. Taking into account that we are considering small deformations, the components $u_i(x_i + \Delta x_i)$ can be expressed in terms of a Taylor series expansion around $x_{i}$ as
\begin{equation}\label{2.0.1}
    u_i(x_i + \Delta x_i) = u_i(x_i) + \sum^3_{j=1} \frac{\partial u_i}{\partial x_j}\Delta x_j + O(\Delta x^2),
\end{equation}
where $O(\Delta x^2)$ are quadratic and higher-order terms, which will not be taken into account. Using equations (\ref{2.0.0}) and (\ref{2.0.1}) the components of the difference between both displacements take the form 
\begin{align}\label{2.1.1}
    \Delta u_{i}(x_{i}) =\sum_{j=1}^{3} \frac{\partial u_{i}}{\partial x_{j}} \Delta x_{j},
\end{align}
where $F_{ij} = \partial u_i/\partial x_j$ are known as the components of the second-order displacement gradient tensor $\mathbf{F}$.

From our tensor calculus courses, a second-order tensor can always be written as the sum of a symmetric $S_{ik}$, and an anti-symmetric $A_{ik}$ part, so that $F_{ik} = S_{ik} + A_{ik}$, where
\begin{align}\label{2.1.2}
    S_{ik} = \frac{1}{2}\left( F_{ik} + F_{ki}\right), \quad
    A_{ik} = \frac{1}{2}\left( F_{ik} - F_{ki}\right).
\end{align}
Therefore, the components of the displacement gradient tensor $\mathbf{F}$ can be rewritten as follows
\begin{equation}\label{2.1.3}
    F_{ij} = \frac{1}{2}\left(\frac{\partial u_{i}}{\partial x_{j}}+\frac{\partial u_{j}}{\partial x_{i}}\right)+\frac{1}{2}\left(\frac{\partial u_{i}}{\partial x_{j}}-\frac{\partial u_{j}}{\partial x_{i}}\right).
\end{equation}
Finally, the strain tensor $e_{ij}$ is defined as the symmetric part of the displacement gradient tensor
\begin{equation}
    e_{i j}=\frac{1}{2}\left(\frac{\partial u_{i}}{\partial x_{j}}+\frac{\partial u_{j}}{\partial x_{i}}\right).\label{eq:ST}
\end{equation}
It is worth mentioning that the anti-symmetric part measures the rate of rotation and for this reason is known as the spin tensor. It quantifies the instantaneous rate of rigid rotation of an infinitesimal volume \cite{Gonzalez2008}.
 
\subsection{Strain and stress relation}

Stress and strain tensors can be related in an elastic media by the constitutive equation 
\begin{equation} \label{2.2.1}
S_{i j}=\sum^{3}_{k=1}\sum^{3}_{l=1} c_{i j k l} e_{k l},
\end{equation}
where $S_{ij}$ are the components of the stress tensor. This equation  corresponds to a generalization of Hooke's law. The proportionality factor, which is known as the elastic tensor, is a fourth-order tensor with 81 components. It is worth saying that of those 81, 21 are independent and are required to specify the stress–strain relationship. However, for isotropic linear elastic materials, the number of independent parameters is reduced to only two \cite{Shearer2019}. In this case, the components of the elastic tensor are defined as
\begin{equation} \label{2.2.2}
c_{i j k l}=\lambda \delta_{i j} \delta_{k l}+\mu\left(\delta_{i l} \delta_{j k}+\delta_{i k} \delta_{j l}\right),
\end{equation}
being $\delta_{ij}$ the components of the Kronecker tensor. The parameters $\lambda$ and $\mu$ are the Lamé parameters of the material and are related with the seismic waves velocities. It is worth mentioning that the isotropy works as a first-order approximation in the interior of the Earth.

Combining equations (\ref{2.2.1}) and (\ref{2.2.2}), we obtain that the components of the stress tensor are given by
\begin{align}\label{2.2.3}
S_{i j} &=\sum^{3}_{k=1}\sum^{3}_{l=1} \left[\lambda \delta_{i j} \delta_{k l}+\mu\left(\delta_{i l} \delta_{j k}+\delta_{i k} \delta_{j l}\right)\right] e_{k l} \nonumber \\ 
&=\lambda \delta_{i j} \sum^{3}_{k=1}\sum^{3}_{l=1} \delta_{k l} e_{k l}+ \mu \sum^{3}_{k=1}\sum^{3}_{l=1}\left(\delta_{i l} \delta_{j k}e_{k l} +\delta_{i k} \delta_{j l}e_{k l}\right) \nonumber\\
&=\lambda \delta_{i j} \sum^{3}_{k=1}e_{k k}+2 \mu e_{i j}.
\end{align}
where $\sum^{3}_{k=1}e_{k k}$ is the trace of the strain tensor. Note that we only need the Lamé parameters in order to describe the linear stress-strain relation within an isotropic solid.

\subsection{Wave propagation}

For an isotropic elastic homogeneous medium, the system of equations for wave propagation can be derived from the Newton’s second law:
\begin{align} \label{2.c1}
\rho \frac{\partial^{2} u_{i}}{\partial t^{2}} = \sum_{j=1}^{3} \frac{\partial S_{ij}}{\partial x_j} + f_{i},
\end{align}
where $\partial S_{ij} / \partial x_j $ denotes the divergence of the stress tensor and $f_{i}$ are force terms, which generally are composed of a gravity term $f_{g}$ and a source term $f_{s}$. However, in regions far from the seismic source and in absence of body forces (\textit{e.g.}, gravity and forces due to electric fields and magnetic fields), this term can be neglected to obtain the homogeneous equation of motion as follows
\begin{align} \label{2.c2}
\rho \frac{\partial^{2} u_{i}}{\partial t^{2}} = \sum_{j=1}^{3} \frac{\partial S_{ij}}{\partial x_j},
\end{align}
where solutions of this equation provide the predicted ground motion at some distance from the source, which is the primary goal of seismic simulations \cite{Shearer2019}.

To solve (\ref{2.c2}), we have to express the stress tensor in terms of the displacement. Fortunate,  substituting (\ref{eq:ST}) into (\ref{2.2.3}) allows to express the components $S_{ij}$ in terms of the displacement components $\mathbf{u}$
\begin{align}\label{2.c3}
S_{i j} = \lambda \delta_{ij} \sum_{k=1}^{3} \frac{\partial u_{k}}{\partial x_k} + \mu \left( \frac{\partial  u_{j}}{\partial x_i} + \frac{\partial  u_{i}}{\partial x_j} \right),
\end{align}
Together with (\ref{2.c2}), equation (\ref{2.c3}) completes the coupled system of equations for the displacement vector and the stress tensor, that can be used to model numerically the wave propagation into a complex media (for more information see page 43 of \cite{Shearer2019}). 

Throughout this work we will focus our attention on P-SV wave propagation in 2D Cartesian coordinates $(x,z)$. Based on this, the components of the displacement vector $\mathbf{u}$ are given by:  $(u_i) = (u_{x},0,u_{z})$. Then, using (\ref{2.c2}) we can write the wave evolution equations for each component of $\mathbf{u}$ as follows:
\begin{align}\label{2.c4}
\rho \frac{\partial^{2} u_x}{\partial t^{2}} = \frac{\partial S_{xx}}{\partial x} + \frac{\partial S_{xz}}{\partial z}, \\
\rho \frac{\partial^{2} u_z}{\partial t^{2}} = \frac{\partial S_{zx}}{\partial x} + \frac{\partial S_{zz}}{\partial z}. 
\end{align}
On the other hand, using (\ref{2.c3}), $S_{xx}$, $S_{xz}$, $S_{zx}$ and $S_{zz}$ can be express in terms of the Lam\'e parameters as
\begin{align}\label{2.c5.1}
S_{xx} &= (\lambda+2 \mu) \frac{\partial u_{x}}{\partial x}+\lambda \frac{\partial u_z}{\partial z},  \\
S_{xz} &= S_{zx} = \mu\left(\frac{\partial u_z}{\partial x}+\frac{\partial u_x}{\partial z}\right), \label{2.c5.2} \\
S_{zz} &= (\lambda+2 \mu) \frac{\partial u_z}{\partial z}+\lambda \frac{\partial u_x}{\partial x}. 
\label{2.c5.3}
\end{align}
Equations (\ref{2.c4})-(\ref{2.c5.3}) constitute the coupled system of equations for two-dimensional P-SV wave propagation that we will solve numerically during this work. However, it is convenient to express these equations as a first-order system of equations by assuming the velocity components as follows $\dot{u}_{i} = \partial u_{i} / \partial t$. Then, we can re-write the PS-V equations as
\begin{align}\label{2.c6.1}
\frac{\partial \dot{u}_{x}}{\partial t} &= \frac{1}{\rho} \left[ \frac{\partial S_{xx}}{\partial x} + \frac{\partial S_{xz}}{\partial z} \right],  \\ \nonumber \\
\frac{\partial \dot{u}_{z}}{\partial t} &= \frac{1}{\rho} \left[ \frac{\partial S_{zx}}{\partial x} + \frac{\partial S_{zz}}{\partial z} \right],  \\ \nonumber \\
\frac{\partial S_{xx}}{\partial t} &= (\lambda+2 \mu) \frac{\partial \dot{u}_{x}}{\partial x} + \lambda \frac{\partial \dot{u}_{z}}{\partial z},  \\ \nonumber \\
\frac{\partial S_{zz}}{\partial t} &= (\lambda+2 \mu) \frac{\partial \dot{u}_{z}}{\partial z} + \lambda \frac{\partial \dot{u}_{x}}{\partial x},  \\ \nonumber \\ \label{2.c6.5}
\frac{\partial S_{xz}}{\partial t} &= \mu \left[ \frac{\partial \dot{u}_{x}}{\partial z} + \frac{\partial \dot{u}_{z}}{\partial x}\right],
\end{align}
where we have taken the first time derivatives of expressions \eqref{2.c5.1}-\eqref{2.c5.3}. Finally, for a complete description of seismic wave propagation, the information about the internal forces and deformations in the material are needed \cite{Shearer2019}. When lateral heterogeneities take place, gradients of the Lamé parameters should appear in the equation of motion \cite{Alterman1970}, and the above system is not longer suitable to deal with this kind of media properties.

\section{Numerical methods}\label{sec:NM}

The system of equations \eqref{2.c6.1}-\eqref{2.c6.5} cannot be solved in an analytical way, especially when dealing with heterogeneous media, mainly because different Lamé parameters need to be considered in the domain. For this reason we developed a code adaptation based on CAFE \cite{Lora2015} and MAGNUS \cite{Navarro2017} codes, which were designed to solve the magneto-hydrodynamics equations in relativistic and Newtonian regimes, respectively. In particular, the adapted code is divided in subroutines. All of them fulfill a different and specific function, some examples of this functions are the storage and recycle of variables, the set up of initial data and boundary conditions, the evolution in time and the data saving.

Before going into the algorithm, a dimensional and characteristic analysis of the equations is needed. In section \ref{sec3A}  a detail description of a dimensional analysis is performed, due to the need to work with different scales. In \ref{sec3B}, through doing a characteristic analysis of the equations, the velocities $v_s$, $v_p$ and two modes that describe the propagation of the waves are found. In section \ref{sec3C} a description of the numerical methods used in the code is presented, adding a brief outline of its operation. 

\subsection{Dimensional analysis}\label{sec3A}

The main purpose of performing a dimensional analysis is to check the dimensional consistency of equations to construct normalized models and then scale up the results \cite{Woodside1972}. In the present case, the main benefit of the dimensional analysis is that it allows us to rewrite the system with dimensionless equations that later can be scaled up. For example, consider a domain of $4$ [km]: It is possible to define a longitude $l_{0}=1$ [km] so that when the domain is divided by it we obtain the non-dimensional quantity $l^{\ast}=4$. The above allows to easily convert the initial domain to a domain of $0$ to $4$, or  $-2$ to $2$, which is clearly dimensionless. Then, it can be easily converted again to a domain of $4$ [km] depending on the units of certain parameters. 

According to the law of dimensional homogeneity, every additive term in an equation has the same dimensions. From this law, it follows that by dividing each term of the equation by a collection of variables and constants, whose product has those same dimensions, the equation turns out to be non-dimensional \cite{RoseHulman}. To eliminate the physical dimensions of the system we define the dimensionless variables $x^{*}$, $z^{*}$, $t^{*}$, $S_{xx}^{*}$, $S_{xz}^{*}$ and $S_{zz}^{*}$ as follows
\begin{align}\label{eq3.1A}
x^{*} &= x/l_{o}, \quad z^{*} = z/l_{o}, \quad t^{*} = t/t_{o}, \nonumber \\ \\
S_{xx}^{*} &= S_{xx}/p_{o},
\quad S_{xz}^{*} = S_{xz}/p_{o},
\quad S_{zz}^{*} = S_{zz}/p_{o}. \nonumber
\end{align}
The above allows rewriting the system of equations as
\begin{align}\label{eq3.2A.1}
\frac{\partial \dot{u}_{x}^{*}}{\partial t^{*}} &= \frac{1}{\rho^{*}} \left[ \frac{\partial S_{xx}^{*}}{\partial x^{*}}+\frac{\partial S_{xz}^{*}}{\partial z^{*}} \right], 
 \\ \nonumber \\ 
\frac{\partial \dot{u}_{z}^{*}}{\partial t^{*}} &=  \frac{1}{\rho^{*}} \left[ \frac{\partial S_{xz}^{*}}{\partial x^{*}}+\frac{\partial S_{zz}^{*}}{\partial z^{*}} \right], 
\\ \nonumber \\ 
\frac{\partial S_{xx}^{*}}{\partial t^{*}} &= (\lambda^{*} + 2\mu^{*}) \frac{\partial \dot{u}_{x}^{*}}{\partial x^{*}} + \lambda^{*} \frac{\partial \dot{u}_{z}^{*}}{\partial z^{*}}, 
\\ \nonumber \\
\frac{\partial S_{zz}^{*}}{\partial t^{*}} &= (\lambda^{*} + 2\mu^{*}) \frac{\partial \dot{u}_{z}^{*}}{\partial z^{*}} + \lambda^{*} \frac{\partial \dot{u}_{x}^{*}}{\partial x^{*}}, 
\\ \nonumber \\ \label{eq3.2A.5}
\frac{\partial S_{xz}^{*}}{\partial t^{*}} &= \mu^{*} \left[ \frac{\partial \dot{u}_{x}^{*}}{\partial z^{*}} + \frac{\partial \dot{u}_{z}^{*}}{\partial x^{*}} \right], 
\end{align}
where all the equations are non-dimensional and $\rho^{*}=\rho/\rho_{o}$, $\lambda^{*}=\lambda/p_{o}$ and $\mu^{*}=\mu/p_{o}$, where $p_{o}$ can be given in terms of density and velocity as $p_{o}=\rho_{o}v_{o}^{2}$.

\subsection{Characteristic analysis}\label{sec3B}

After doing the dimensional analysis it follows studying the characteristic structure of our system. This part allows us to determine the characteristic velocities, in this case: the velocities of P- and SV- waves. Moreover the characteristic analysis will allow to impose the boundary conditions discussed later.  

The first step is to write the equations in a compact matrix form. In this case the system of PDEs \eqref{2.c6.1}-\eqref{2.c6.5} can be written as
\begin{align}\label{eq3.1B}
\frac{\partial \mathbf{U}}{\partial_{t}} + \mathbf{A}_{x} \frac{\partial \mathbf{U}}{\partial_{x}} + \mathbf{A}_{z} \frac{\partial \mathbf{U}}{\partial_{z}} = 0,
\end{align}
with 
\begin{align}\label{eq3.2B}
\mathbf{A}_{x} =
\begin{bmatrix}
0 & 0 &  -1/\rho & 0 & 0 \\ 
0 & 0 &  0 & -1/\rho & 0 \\ 
-(\lambda + 2\mu) & 0 & 0 & 0 & 0 \\ 
0 & -\mu & 0 & 0 & 0 \\ 
-\lambda & 0 & 0 & 0 & 0 
\end{bmatrix},
\end{align}
\begin{align}\label{eq3.3B}
\mathbf{A}_{z} =
\begin{bmatrix}
0 & 0 & 0 & -1/\rho & 0 \\ 
0 & 0 &  0 & 0 & -1/\rho \\ 
0 & -\lambda & 0 & 0 & 0 \\ 
-\mu & 0 & 0 & 0 & 0 \\ 
0 & -(\lambda + 2\mu) & 0 & 0 & 0 
\end{bmatrix},
\end{align}
and the vector $\mathbf{U}$ defined as
\begin{align}\label{eq3.4B}
\mathbf{U} = [ v_{x} \quad v_{z} \quad S_{xx} \quad S_{xz} \quad S_{zz} ]^{T}. 
\end{align}
Now, in order to get the characteristic structure, it is necessary to calculate the eigenvalues of both matrices $\mathbf{A}_{x}$ and $\mathbf{A}_{z}$. The respective eigenvalues of the matrix along the x-direction are:
\begin{align}\label{eq3.5B}
\lambda_{1} = 0, \quad \lambda_{2,3} = \pm \sqrt{\frac{\mu}{\rho}}, 
\quad \lambda_{4,5} = \pm \sqrt{\frac{\lambda + 2\mu}{\rho}},
\end{align}
and determine the characteristic velocities of the system, where $v_{p}$ corresponds to compressional waves, while $v_{s}$ to shear waves
\begin{align}\label{eq3.6B}
v_{p} = \sqrt{\frac{\lambda + 2\mu}{\rho}}, \qquad v_{s} = \sqrt{\frac{\mu}{\rho}}.
\end{align}

Because compressional waves travel faster and arrive first, they are called primary or P-waves, whereas the later arriving shear waves are called secondary or SV-waves. If velocities $v_{s}$ and $v_{p}$ are only a function of depth, as the present case, the material can be modelled as a series of homogeneous layers with no gradients in the Lam\'e parameters within each layer \cite{Shearer2019}. More importantly, as $v_{s}$ only depends on $\mu$, SV-wave propagation is pure shear with no volume change, whereas $P$-wave propagation involves shearing as well as compression in the material. 

The analysis of eigenvalues and eigenvectors of the system of equations allows rewriting the matrices in terms of the wave velocities. Moreover, $\mathbf{A}_{x}$ can be written in terms of the eigenvectors matrix $\mathbf{P}_{x}$ and the eigenvalues matrix $\mathbf{D}$ thought the relation $\mathbf{A}_{x} = \mathbf{P}_{x} \mathbf{D} \mathbf{P}^{-1}_{x}$. The same procedure can be applied to $\mathbf{A}_{z}$. As a consequence, equation (\ref{eq3.1B}) can be written for the x-component as follows
\begin{align}\label{eq3.7B}
&\partial_{t} \mathbf{W}_{x} + \mathbf{D} \partial_{x} \mathbf{W}_{x} = 0, \nonumber \\ \nonumber \\
&\mathbf{W}_{x}
= \mathbf{P}^{-1}_{x} \mathbf{U}
= \frac{1}{2}
\begin{bmatrix}
2S_{zz} - 2\alpha S_{xx}\\ 
\alpha S_{xx} + \alpha v_{p} \rho v_{x}\\ 
\alpha S_{xx} - \alpha v_{p} \rho v_{x}\\ 
S_{xz} + v_{s} \rho v_{z}\\ 
S_{xz} - v_{s} \rho v_{z}
\end{bmatrix},
\end{align}
where
\begin{align}\label{eq3.8B}
\alpha = \frac{v_{p}^{2}-2v_{s}^{2}}{v_{p}^{2}}.
\end{align}
The vector $\mathbf{W}_{x}$ allows us to identify fundamental modes that propagate to left and right. For this reason, we can write the vector $\mathbf{W}_{x}$ as 
\begin{align}\label{eq3.9B}
\mathbf{W}_{x} = [ S_{zz} - \alpha S_{xx} \quad M_{1L} \quad M_{1R} \quad M_{2L} \quad M_{2R} ]^{T},
\end{align}
where $M_{1L}$ and $M_{2L}$ are modes that propagate to the left, and $M_{1R}$ and $M_{2R}$ modes that propagate to the right. The same procedure can be done for the z-component obtaining two modes that propagate both up and down. The above will be very useful when defining the boundary conditions.

\subsection{Code description}\label{sec3C}

At this point it is clear that the first step in the analysis of a physical system is to formulate a mathematical model as a system of equations. Once this is done, the next task that is usually  challenging is to solve the equations. Our seismic equations for P-SV wave propagation are solved numerically using a FORTRAN program equipped with numerical grid generation, initial conditions, boundary conditions, spatial discretization, and time integration. 

\begin{figure}
\centering
\includegraphics[scale=0.85]{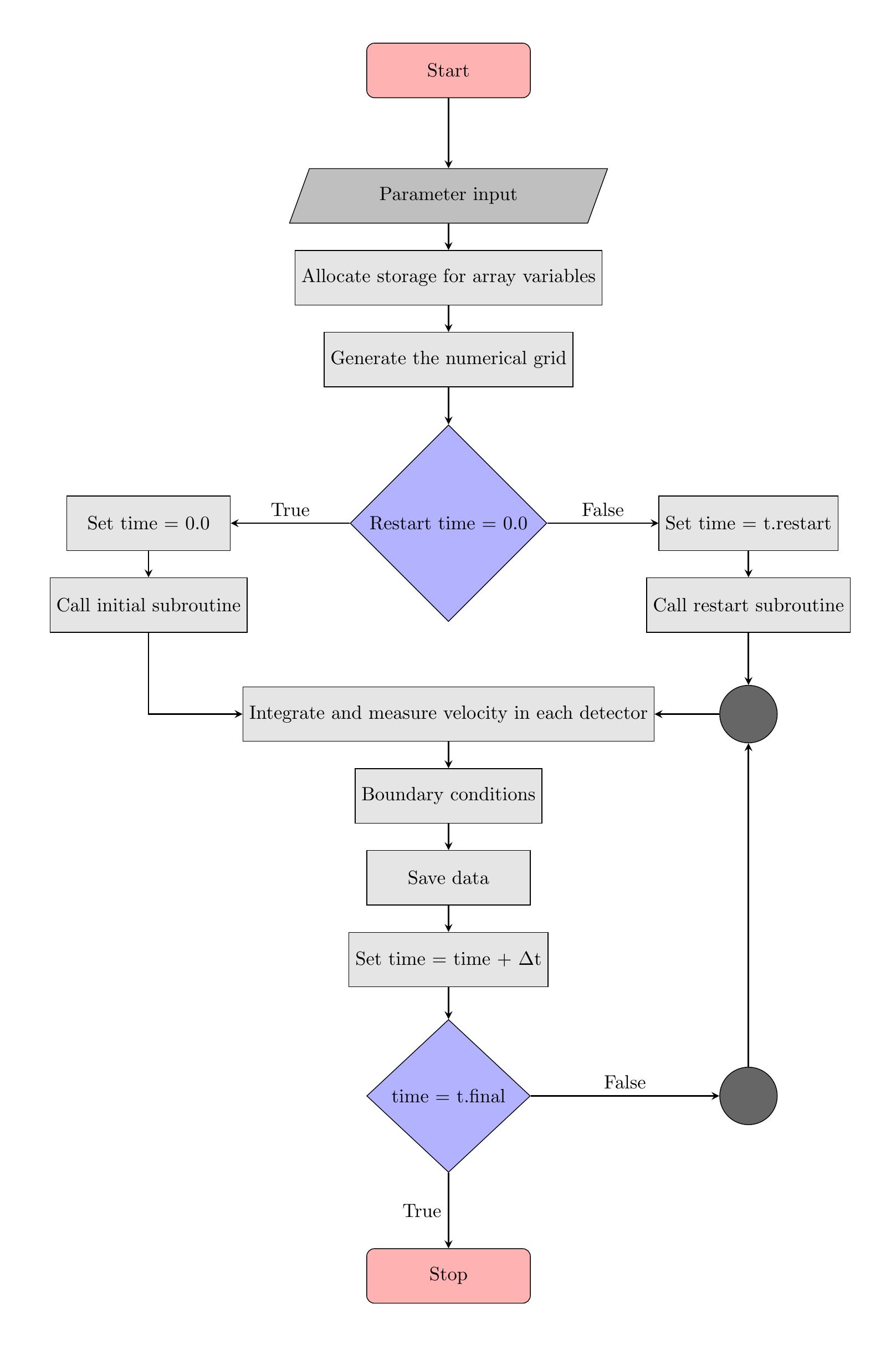}
\caption{Graphical representation of the algorithm.}
\label{flowchart}
\end{figure}

Fig.\ref{flowchart} shows a flowchart of the algorithm, that is, of the set of instructions and processes that the code follows and carries out to solve the problem. As the figure shows, the first step in the code execution is the parameter input. In this step is necessary to introduce all the parameters that the program requires to solve the system. Some of them are the domain size and the number of points of the grid, with them the code calculates the spatial and temporal resolution and generates the numerical grid. Once this is done, the code initializes all the arrays and calls an initial subroutine. Here, parameters like $\rho$, $\lambda$ and $\mu$ are defined for each zone in the domain. Also here, the detectors are located in different grid points taking into account the position of the first detector and the spacing between the consecutive ones. When time starts to evolve the numerical integration begins. Reviewing the boundary conditions imposed, the data obtained is organized and saved in each time step until the time reaches its final value. It should be clarified that the parameter \textit{t.restart} is introduced only in case that the code execution is aborted (for example, when the execution of the code ends and the simulation time was not enough to observe the expected phenomena). In that case, it is not necessary to start the execution from $t=0.0$, but from the end time of the previous execution which must be defined as the restart time. 

\subsubsection*{\textbf{Method of lines}}

Taking into account all the above, now it is time to address the issue of numerical integration. In many numerical analysis, the system is solved using schemes that discretize the system simultaneously in both space and time (see \cite{Virieux1986, Vossen2002}). However, our numerical solution is implemented through the method of lines. This method is a well-established numerical procedure that transforms PDEs into ordinary differential equations (ODEs).

The fundamental idea of the method of lines is to replace the spatial derivatives with algebraic approximations \cite{Schiesser2009}. Once this is done only the time remains as an independent variable and we obtain a system of ordinary equations that approximate the original system. In other words, the method of lines allows to rewrite the original system of PDEs as a system of ODEs at each node of the numerical grid. At each of these nodes the derivatives are numerically approximated using forward, backward, and central finite-differences. Therefore, the main problem consists in formulating an approximate system of ODEs and then apply an integration method to compute the numerical solution of the problem. In particular, we use a second-order finite-difference scheme couple to a fourth-order Runge-Kutta time integrator. 

\subsubsection*{\textbf{Finite-Differences and Time Step}}

The finite-difference approximation allows us to define the variables in a discrete set of points $(x_{i},t^{n})$ that constitutes the grid where we seek to find the solution \cite{Thomas2013}. Thus, any analytical function $f(x,t)$ is defined only at each point of the numerical grid. We can denote the value of the function in some point $x_{i}$ during a time $t^{n}$ as $f_{i}^{n}$. The set of points $(x_{i},t_{n})$ is given by $(i \Delta x, n \Delta t)$, where $\Delta x$ and $\Delta t$ define the spatial and temporal resolution. Clearly, $i=0,1,\cdots,N_{x}$ and $n=0,1,\cdots,N_{t}$, where $N_{x}$ is the number of grid points in the x direction and $N_t$ the number of time levels. Once $N_x$ defined as an input parameter, it can be used in the code to determine the spatial resolution as
\begin{align}\label{eq3.1C}
\Delta x = \frac{x_{max}-x_{min}}{N_{x}},
\end{align}
where $x_{min}$ and $x_{man}$ are the left and right limits of the domain. The numerical grid is constructed by defining each $x_{i}$ with the following expression
\begin{align}\label{eq3.2C}
x_{i} = x_{min} + i\Delta x. 
\end{align}
For greater values of $N_{x}$ the number of points $x_{i}$ is also greater, while their separation $\Delta x$ becomes smaller increasing the spatial resolution.

For a given time, the function $f(x)$ can be approximated in each point $x_{i}$ by using truncated Taylor series expansions of $f$ around $f_{i}$ \cite{Guzman2010}. For a second-order finite-difference scheme we have
\begin{align}\label{eq3.3C}
f_{i+1} = f_{i} + \Delta x f'_{i} + \frac{\Delta x^{2}}{2} f''_{i} + \Delta(O^{3}), \\
f_{i-1} = f_{i} - \Delta x f'_{i} + \frac{\Delta x^{2}}{2} f''_{i} - \Delta(O^{3}),
\end{align}
where $\Delta(O^{3})$ is the truncation error, which includes cubic and higher-order terms. It is possible to find an expression for the first derivative operator with a second-order error, by combining these last expressions,
\begin{align}\label{eq3.4C}
f'_{i} = \frac{f_{i+1}-f_{i-1}}{2\Delta x} + \Delta(O^{2}).
\end{align}
This expression is known as a central finite difference scheme because it is necessary to know the values of $f_{i+1}$ and $f_{i-1}$ to calculate $f'_{i}$. Therefore, it only works for the points located inside the grid and not in the boundary.

For the points located at extremes of the domain, we use the forward and backward finite-differences. In particular, we use
\begin{align}\label{eq3.5C}
f_{i+1} = f_{i} + \Delta x f'_{i} + \frac{\Delta x^{2}}{2} f''_{i} + \Delta(O^{3}), \\
f_{i+2} = f_{i} + 2\Delta x f'_{i} + \frac{4\Delta x^{2}}{2} f''_{i} + \Delta(O^{3}),
\end{align}
to obtain the forward difference that allows us to calculate the function at $x=x_{min}$
\begin{align}\label{eq3.6C}
f'_{0} = \frac{-f_{i+2} + 4f_{i+1} - 3f_{i}}{2\Delta x} + \Delta(O^{2}).
\end{align}
The same procedure can be applied using $f_{i-1}$ and $f_{i-2}$ to obtain the backward difference that allows us to calculate the function at $x=x_{max}$ 
\begin{align}
f'_{N_{x}}=\frac{f_{i-2} - 4f_{i-1} + 3f_{i}}{2\Delta x} + \Delta(O^{2}).
\end{align}
Of course for $i=0$ and $i=N_x$ it is necessary to implement some boundary conditions, which will be discussed later. Once the space discretization is done, follows the time integration. In this second part we use the method of lines coupled with a fourth-order Runge-Kutta time integrator.

A procedure used to verify the stability of finite-difference schemes applied to linear partial differential equations corresponds to the Von Neumann stability analysis \cite{Thomas2013}. The analysis is based on the Fourier decomposition of the solution of the system of equations \eqref{2.c6.1}-\eqref{2.c6.5}. To do this, we assume the solution in the slice of time $n$ as follows
\begin{equation}
f^{n}_i = \xi^{n}(\Delta x, \Delta t, k) e^{i k x_i},\label{eq:VN}
\end{equation}
where $k$ is the spatial wave number and $\xi^{n}(\Delta x, \Delta t, k)$ is a complex number called the amplification factor. This stability criterion states that the modulus of the amplification factor is always less or equal than $1$, i.e., $\xi \xi^{*} \leq 1$. For our case, this criterion reduces to that the Courant-Friedricks-Levy parameter,  $C_{CFL}$, must be less than or equal to 1, i.e., that $v \Delta t / \Delta x \leq 1$. The Courant-Friedricks-Levy condition states that the distance that a wave travels during a time-step length within the mesh must be lower than the distance between grid elements. Furthermore, the $C_{CFL}$ number defines an upper limit on $\Delta t$ for any given $v$ and $\Delta x$ \cite{Schiesser2009}. Following the Courant–Friedrichs–Levy condition, the time step is chosen as
\begin{align}
\Delta t = \frac{C_{CFL} \cdot min(\Delta x, \Delta z)}{v_{max}},
\end{align}
where we have taken the value $C_{CFL}=0.2$. It is important to clarify that the only criteria to set the value of the $C_{CFL}$ is the Courant-Friedricks-Levy condition. In this case, we chose a small $C_{CFL}$ to guarantee a stable run and not lose information between consecutive time steps (if a larger value is chosen, like $0.5$ or $0.6$, the time step also becomes larger), but it's possible to choose any value less than or equal to $1$. Finally, it is worth mentioning that $v_{max}$ depends on the values of wave velocities $v_{s}$ and $v_{p}$, i.e., $v_{max}=max(v_{s},v_{p})$.

\section{Initial data and boundary conditions}\label{sec:IDBC}

\begin{figure}
\begin{center}
\includegraphics[scale=0.6]{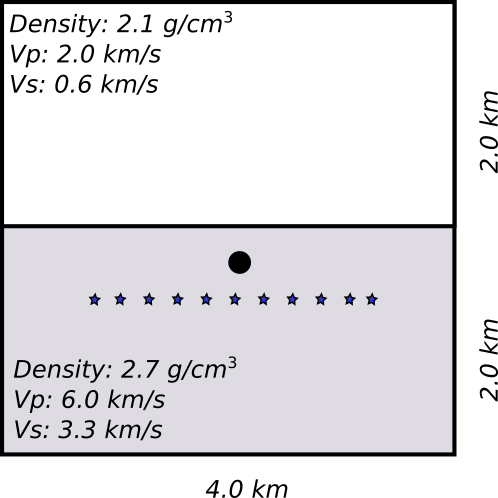}
\caption{Domain scheme for the simulation. Density profile, P-wave and SV-wave velocities for both layers are specified in the figure. The black dot represents the source position at $x=0.0$ and $z=-0.2$ for both homogeneous (without interface and only the lower layer parameters in the entire domain) and heterogeneous media (with interface and the two-layered media). The blue stars represent the positions of eleven equally-spaced detectors, located between $x=-1.0$ $x=1.0$, that measure the z-velocity.}
\label{fig0}
\end{center}
\end{figure}

For the simulation we placed an interface splitting the domain in equal layers as can be seen in Fig.\ref{fig0}. Each layer has a thickness of $2.0$ [km]. The lower layer is a solid with density $2.7$ [g/cm$^{3}$] and P- and SV- wave velocities of $6.0$ [km/s] and $3.3$ [km/s] respectively. The upper layer, which also represent a solid phase, has values of density $2.1$ [g/cm$^{3}$] and P- and SV- wave velocities of $2.0$ [km/s] and $0.6$ [km/s] respectively. This parameters simulate the interface between limestone and wet sand.

Limestone is one of the many types of sedimentary rocks that are generated by sedimentary processes. It is composed mostly of calcite, a mineral whose main component is calcium carbonate. Due to the decrease in calcium carbonate saturation with depth, the production of this chemical compound is limited to shallow water areas of the ocean \cite{Boggs2014}. As a result, the sediment in shallow water marine environments is often composed of limestone. There, the limestone is in contact with other sedimentary rocks, coral reefs and of course wet sand. For this reason, the chosen parameters simulate a real environment that takes place in shallow water. In particular, we set each layer with a thickness of $2.0$ [km], however, it must be mentioned that the thickness, as well as the other media parameters, of each layer is an arbitrary decision of the student and can be adjusted to simulate more realistic environments.

There exist many time-depending sources used in numerical simulation of seismic waves. However, Ricker wavelet has proven to be one of the best tools for processing seismic data, due to it leads to better approximations \cite{Ricker2014}. In order to compare with more than one source results, a sine wavelet is also include. These functions are given by \cite{Cao1992,Wang2015}:
\begin{gather}\label{eq3.3a}
s_{1}(t) = 1000 \left( \sin{(2\pi\omega t)} - \frac{1}{2}\sin{(4\pi\omega t)}\right),  
\\ \nonumber \\ \label{eq3.3b}
s_{2}(t) = 1000\left(1-\frac{1}{2} \omega^{2} t^{2} \right)  \exp \left( -\frac{1}{4} \omega^{2} t^{2} \right), 
\end{gather}
where $\omega$ is the angular frequency of the sine wavelet, whose value is $8.0$ [rad/s], and in the case of the Ricker wavelet is a parameter to control the amplitude of the pulse and the wavelength content. In this last case, the value used in the simulations is $\omega = 50$ [Hz]. Due to the characteristics of the problem, where the wave propagates in two dimensions, the source corresponds to a line, invariant under translations along $y$ direction, which is commonly used \cite{Virieux1986, Vossen2002}. Finally, the sources are added to the equations responsible of the evolution of the normal stresses ($S_{xx}$ and $S_{zz}$) during a period of time.

On the other hand, one of the most used boundary conditions in numerical modelling is the outgoing wave conditions. It is an appropriate boundary condition to represent the radiative phenomena of an isolated system, which generates signals that propagate outward. In the case of the wave equation, for instance, we think of boundary conditions that allow the wave to leave the finite domain. In this case, the main concern is to avoid that the information, which is supposed to leave the domain, is reflected causing interference with the values of the functions that are being calculated within the numerical domain. 

In this work we follow the same procedure as in \cite{Guzman2010}, in which the characteristic structure is used to impose the outgoing boundary conditions. This method consists of eliminating the modes that travel to the opposite boundary. For example, the  condition in the right boundary consists of eliminating the modes that travel to the left and the condition in the left boundary consists of eliminating the modes that travel to the right. With these boundary conditions the reflection into the domain is not allowed.

In order to illustrate the method, consider the vector $\mathbf{W}_x$ in equation (\ref{eq3.7B}), that is 
\begin{align}\label{eq:BC2}
\mathbf{W}_{x}
= \frac{1}{2}
\begin{bmatrix}
2S_{zz} - 2\alpha S_{xx}\\ 
\alpha S_{xx} + \alpha v_{p} \rho v_{x}\\ 
\alpha S_{xx} - \alpha v_{p} \rho v_{x}\\ 
S_{xz} + v_{s} \rho v_{z}\\ 
S_{xz} - v_{s} \rho v_{z}
\end{bmatrix}
= 
\begin{bmatrix}
S_{zz} - \alpha S_{xx}\\ 
M_{1L} \\ 
M_{1R}\\ 
M_{2L}\\ 
M_{2R}
\end{bmatrix}.
\end{align}
This vector, whose components are the characteristic variables, allows to detached the system of equations \eqref{2.c6.1}-\eqref{2.c6.5} in such away the dynamics of the state variables is decomposed into fundamental modes that propagate to the left ($M_{1L}, M_{2L}$) and right ($M_{1R}, M_{2R}$). Explicitly in the right boundary $x=x_{N}$ is required that 
\begin{align}
    \nonumber \frac{1}{2}\left[ \alpha S_{xx}\rvert_{x_N} - \alpha v_{p} \rho v_{x}\rvert_{x_N} \right] &= M_{1R}\rvert_{x_N}, \\ 
    \frac{1}{2}\left[ \alpha S_{xx}\rvert_{x_N} + \alpha v_{p} \rho v_{x}\rvert_{x_N} \right] &= M_{1L}\rvert_{x_N} = 0, \\ \nonumber 
    \frac{1}{2}\left[ S_{xz}\rvert_{x_N} - v_{s} \rho v_{z}\rvert_{x_N} \right] &= M_{2R}\rvert_{x_N}, \\ 
    \frac{1}{2}\left[ S_{xz}\rvert_{x_N} + v_{s} \rho v_{z}\rvert_{x_N} \right] &= M_{2L}\rvert_{x_N} = 0,
\end{align}
while in the left one $x=x_{0}$ 
\begin{align}
    \nonumber \frac{1}{2}\left[ \alpha S_{xx}\rvert_{x_0} - \alpha v_{p} \rho v_{x}\rvert_{x_0} \right] &= M_{1R}\rvert_{x_0} = 0,  \\ 
    \frac{1}{2}\left[ \alpha S_{xx}\rvert_{x_0} + \alpha v_{p} \rho v_{x}\rvert_{x_0} \right] &= M_{1L}\rvert_{x_0}, \\ \nonumber
    \frac{1}{2}\left[ S_{xz}\rvert_{x_0} - v_{s} \rho v_{z}\rvert_{x_0} \right] &= M_{2R}\rvert_{x_0} = 0, \\ 
    \frac{1}{2}\left[ S_{xz}\rvert_{x_0} + v_{s} \rho v_{z}\rvert_{x_0} \right] &= M_{2L}\rvert_{x_0}.
\end{align}
Solving each system of equation on the respective boundary, we obtain 
\begin{align}
   \alpha S_{xx}\rvert_{x_N} &= - \alpha v_{p} \rho v_{x}\rvert_{x_N} = M_{1R}\rvert_{x_N}, \\
   S_{xz}\rvert_{x_N} &= - v_{s} \rho v_{z}\rvert_{x_N} = M_{2R}\rvert_{x_N}, \\
     \alpha S_{xx}\rvert_{x_0} &= \alpha v_{p} \rho v_{x}\rvert_{x_0} = M_{1L}\rvert_{x_0},  \\ 
    S_{xz}\rvert_{x_0} &=  v_{s} \rho v_{z}\rvert_{x_0} = M_{2L}\rvert_{x_0}.
\end{align}
Now the problem has been reduced to compute the values $M_{1R}\rvert_{x_N}$, $M_{1L}\rvert_{x_N}$, $M_{2R}\rvert_{x_N}$, $M_{2L}\rvert_{x_N}$, $M_{1R}\rvert_{x_0}$, $M_{1L}\rvert_{x_0}$, $M_{2R}\rvert_{x_0}$, $M_{2L}\rvert_{x_0}$. To know these values it is enough to make an extrapolation using the internal points of the mesh, that is 
\begin{align}
    f_0 &= 3 f_1 - 3 f_2 + f_3, \\
    f_{Nx} &= 3 f_{Nx - 1} - 3 f_{Nx - 2} + f_{Nx - 3}, 
\end{align}
being $f$ any of the propagation modes. Finally, an analogous procedure is done for the vector $\mathbf{W}_z$, which is associated with the matrix $\mathbf{A}_z$.

\section{Numerical Experiments}\label{sec:results}

For our numerical experiments, unless otherwise stated, all the results corresponding to the solutions of the system of equations \eqref{eq3.2A.1}-\eqref{eq3.2A.5} are obtained in the simulation box $(-2,-2) \le (x,z) \le (2,2)$, where $x$ is the horizontal and $z$ the vertical coordinates. The basic equations are discretized on an uniform cartesian grid with $(N_x \times N_z) = (1600 \times 1600)$ grid points and Courant-Friedrichs-Levy parameter equal to $0.2$. 

Particularly, we carried out four different simulations using the time depending sources mentioned before: a sine wavelet and a Ricker wavelet (see equations \eqref{eq3.3a} and  \eqref{eq3.3b}), each one in both homogeneous and heterogeneous media. In this section we compare the velocity maps of the four numerical simulations at times $t=0.2$ [s] and $t=0.3$ [s]. Furthermore, we also compare the z-velocity as a function of time measured in different detectors distributed along the numerical domain and evaluated in a the time window $t \in [0,0.3]$ [s]. 

\begin{figure}
\begin{center}
\includegraphics[scale=0.42]{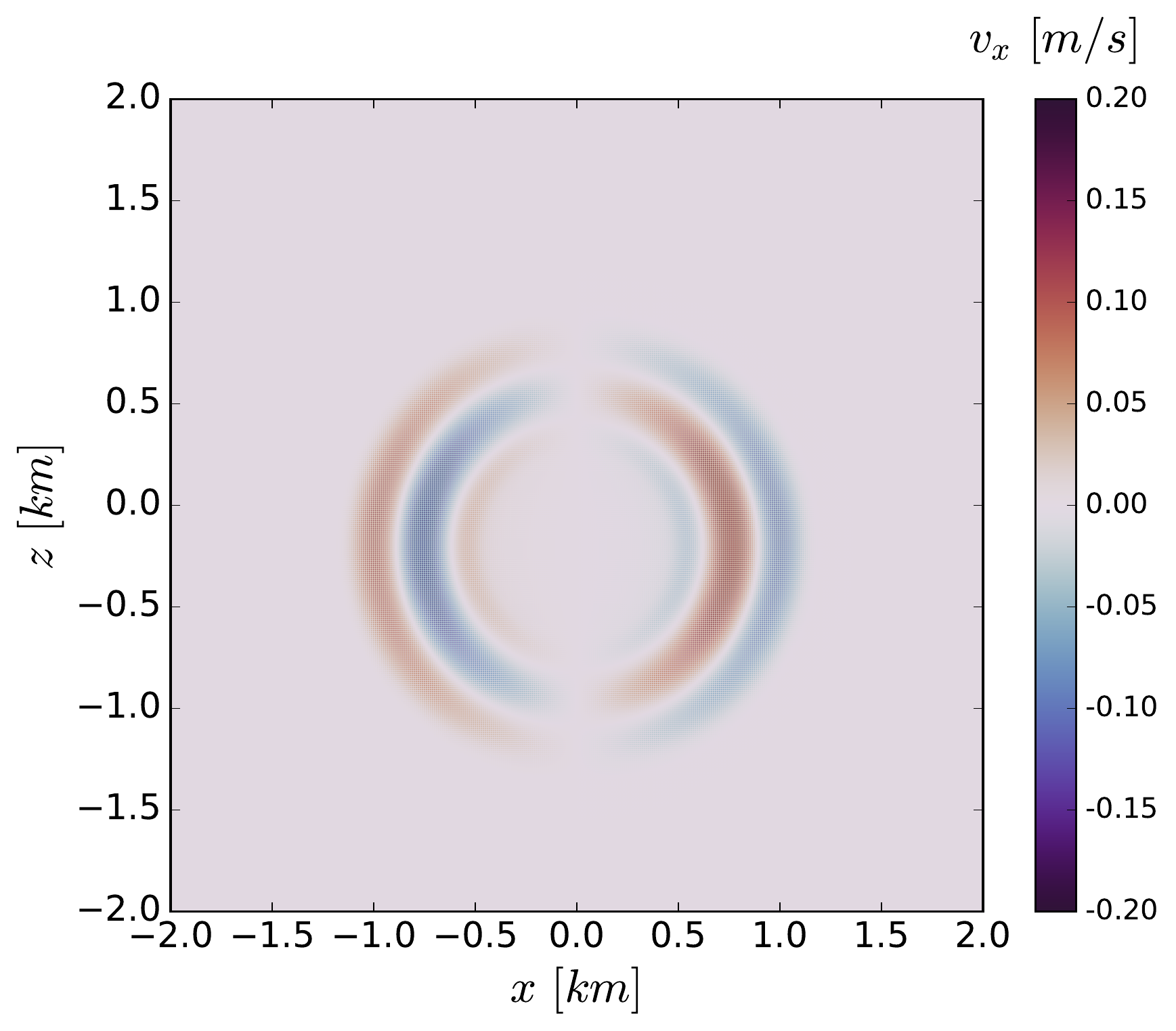} \includegraphics[scale=0.42]{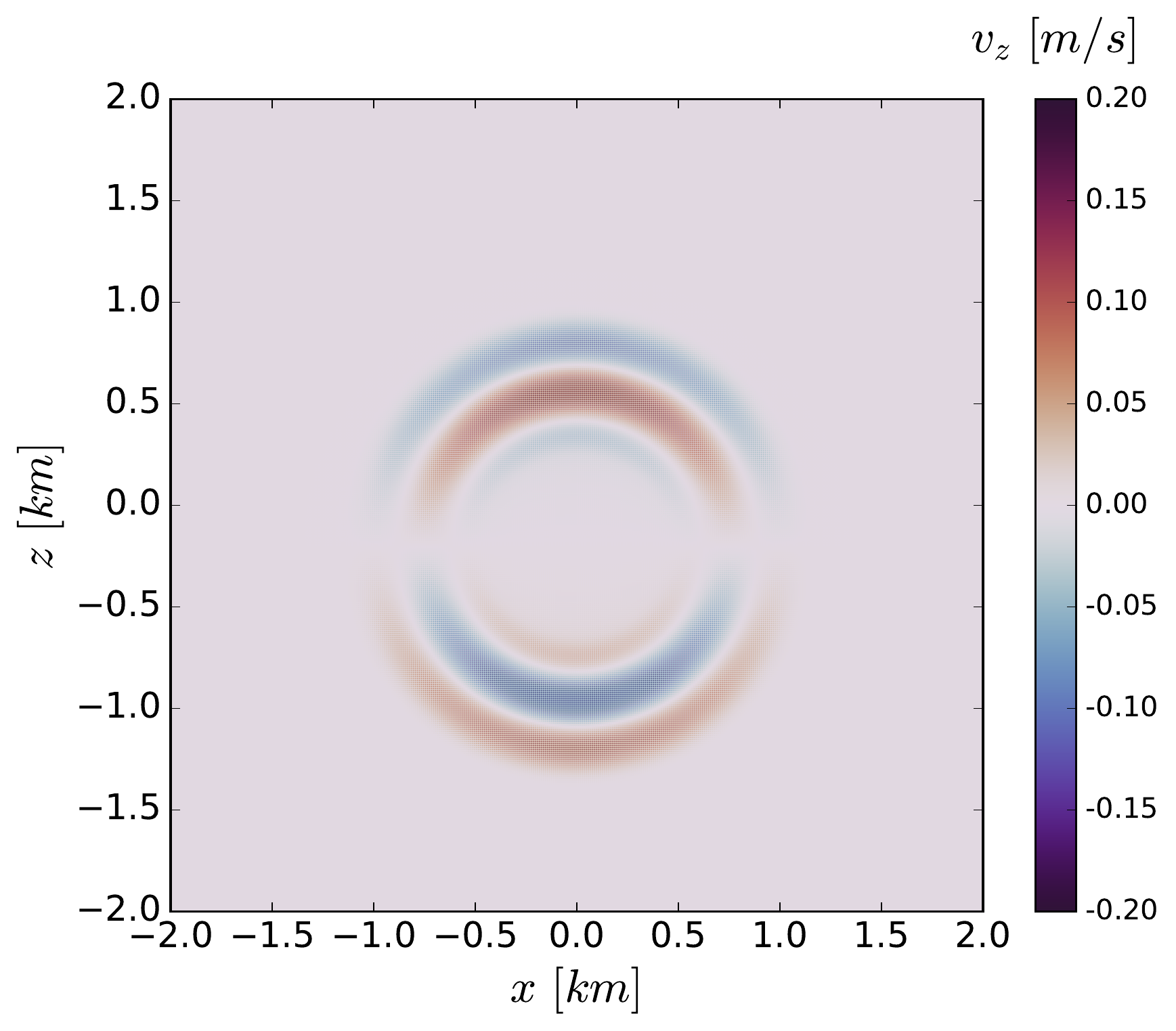}
\includegraphics[scale=0.42]{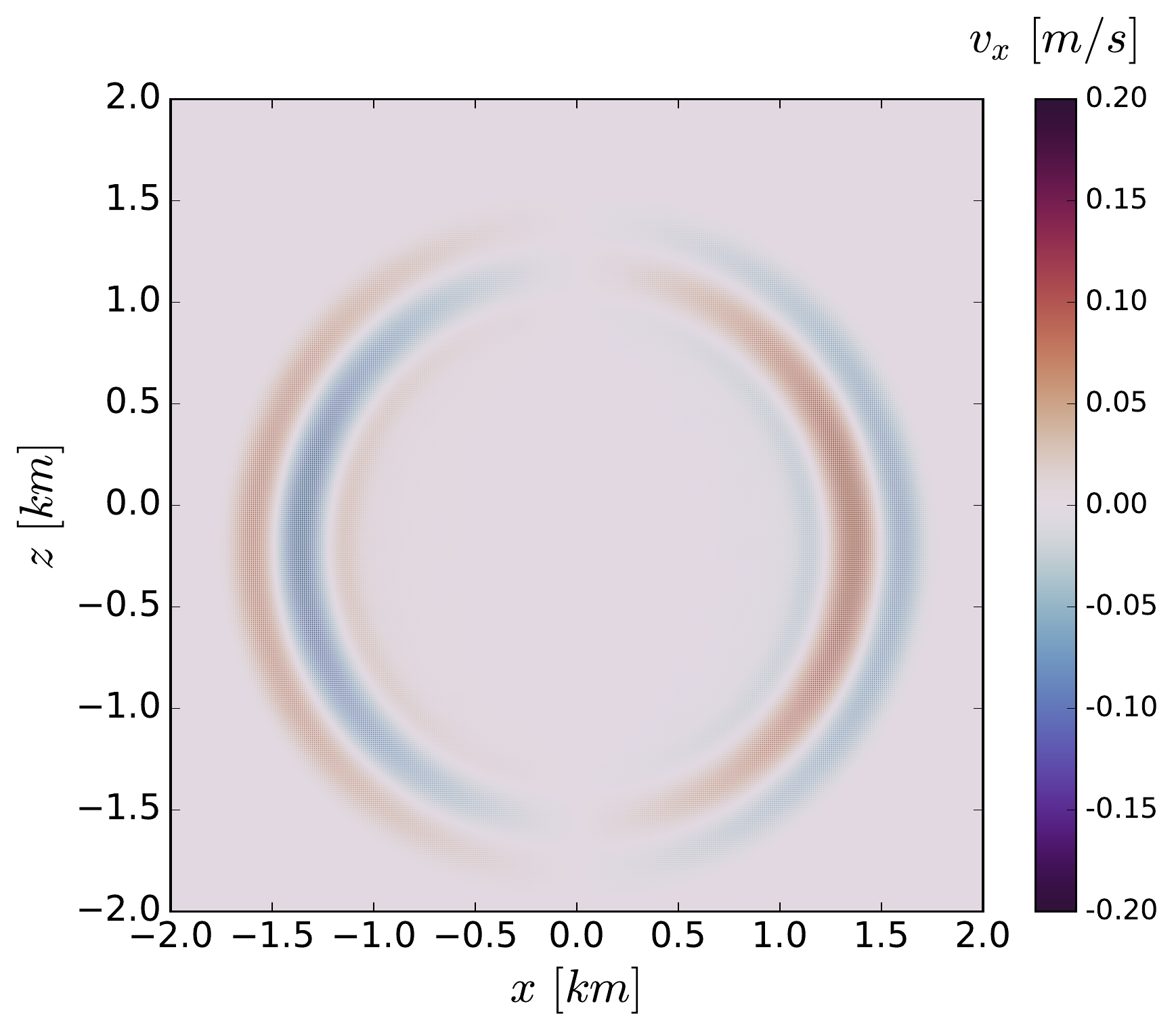} \includegraphics[scale=0.42]{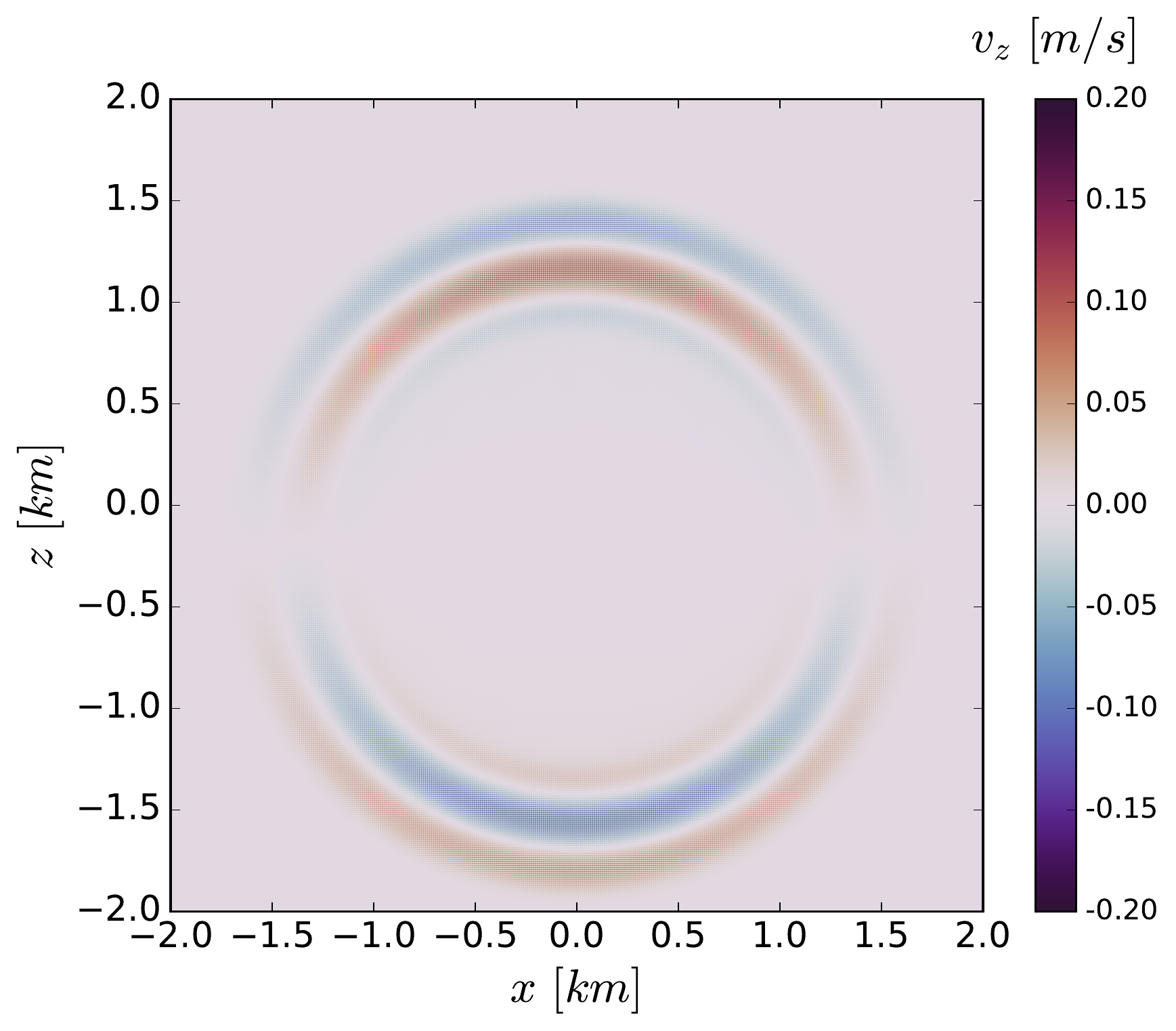}
\caption{Velocity maps for the sine wavelet source applied to the normal stresses in homogeneous media. Top-left and top-right figures represent the horizontal and vertical components of the velocity at time $t=0.2$ [s], while bottom-left and bottom-right figures at time $t=0.3$ [s].}
\label{fig1}
\end{center}
\end{figure}

\begin{figure}
\begin{center}
\includegraphics[scale=0.42]{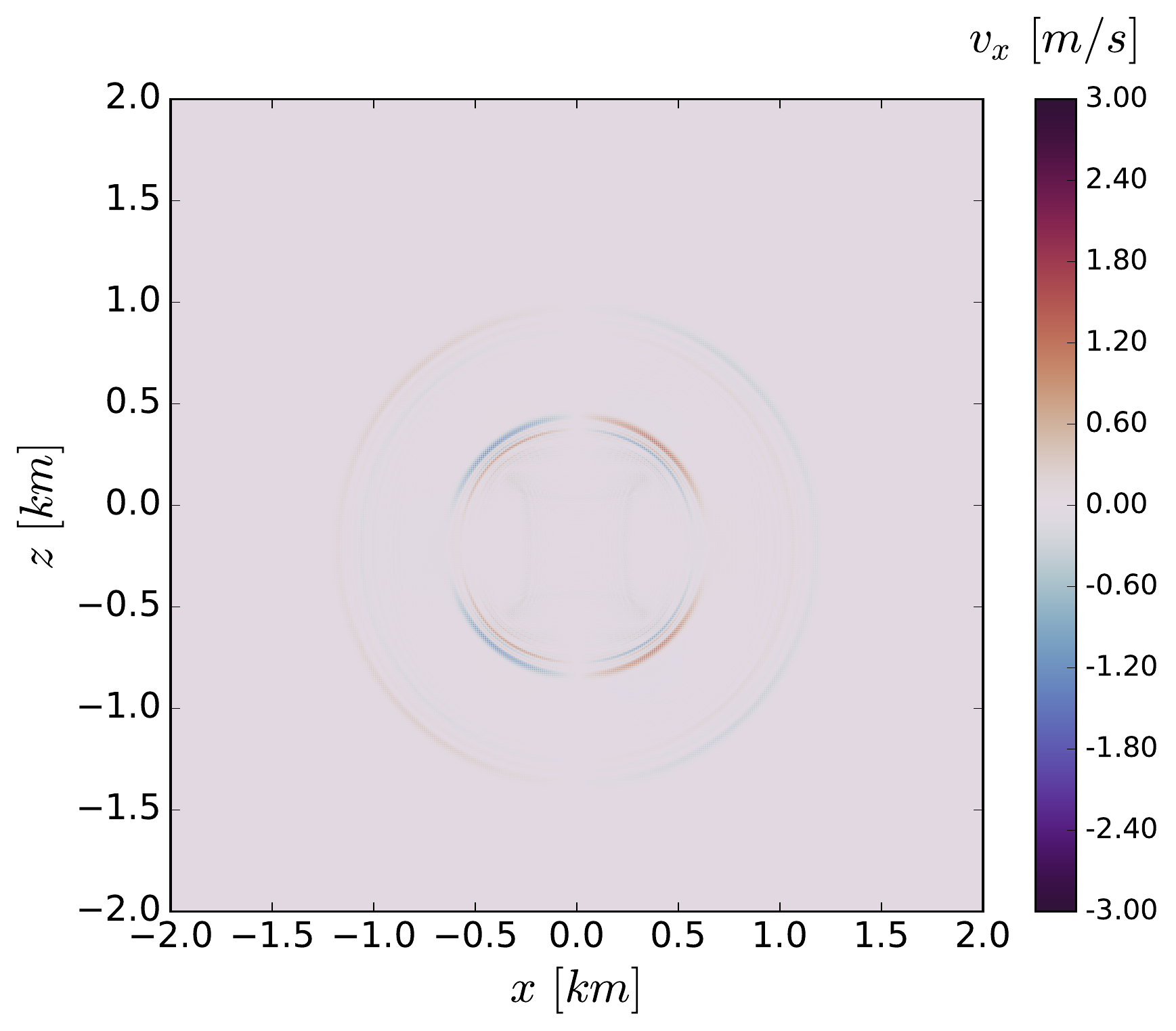} \includegraphics[scale=0.42]{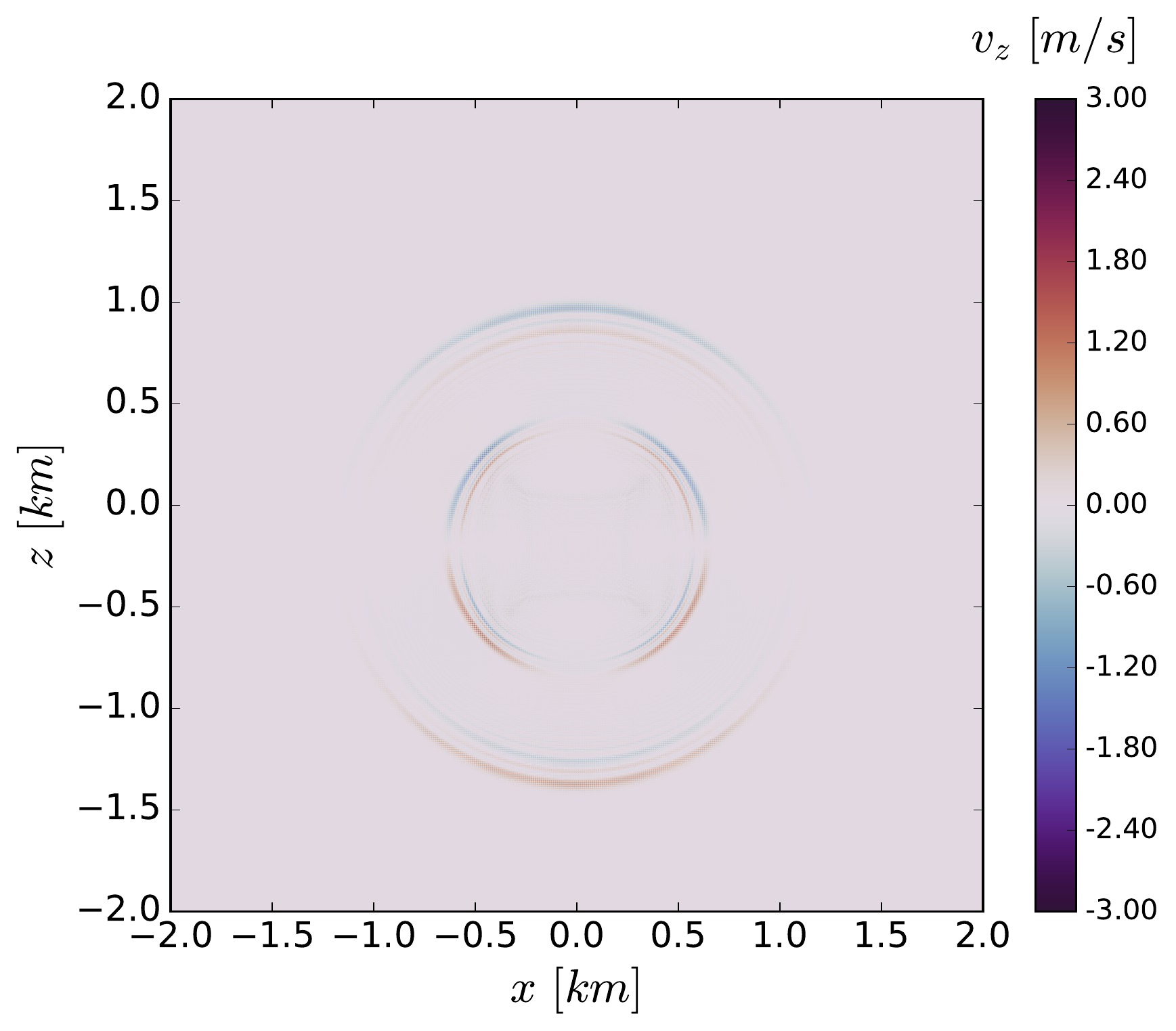}
\includegraphics[scale=0.42]{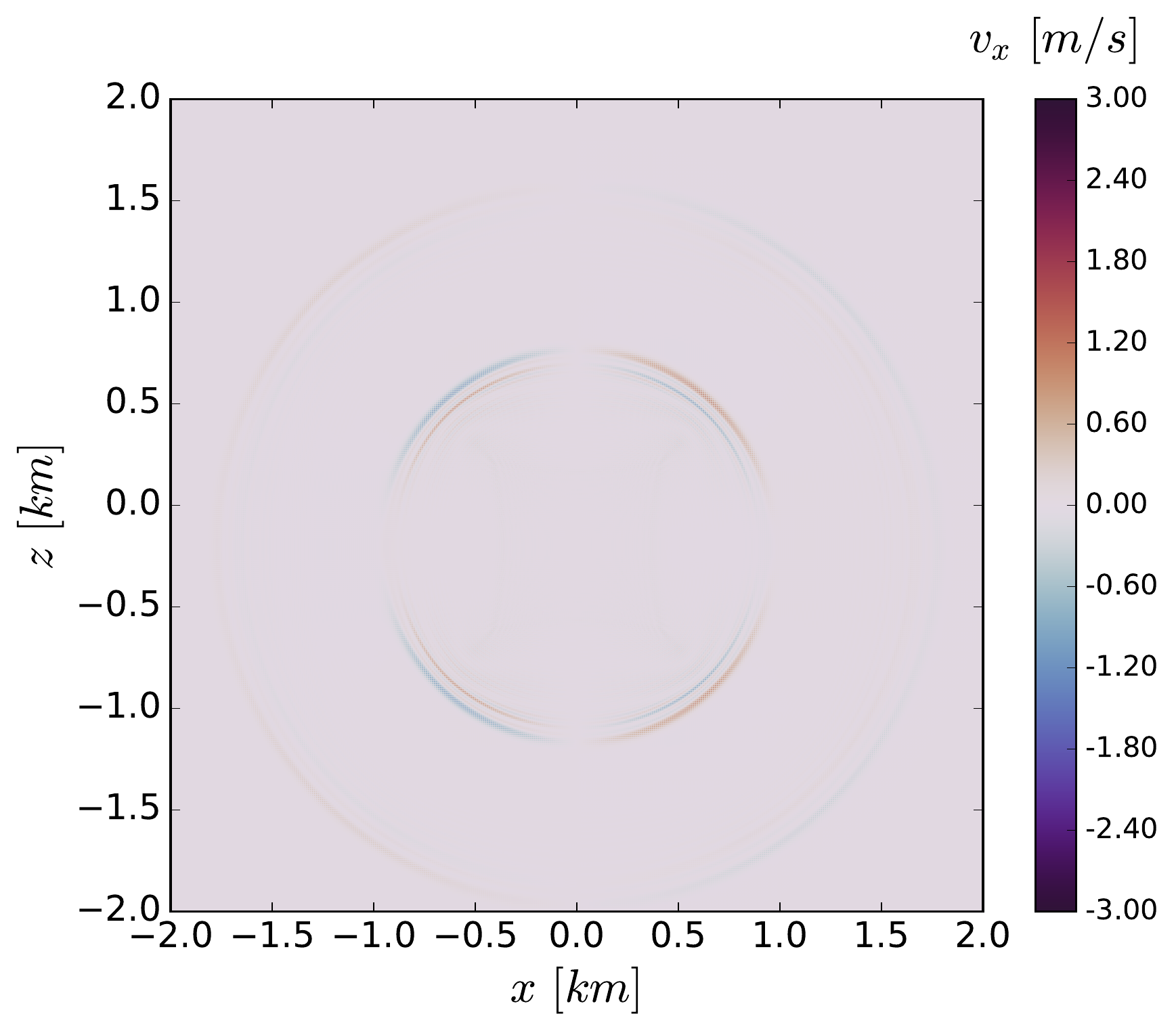} \includegraphics[scale=0.42]{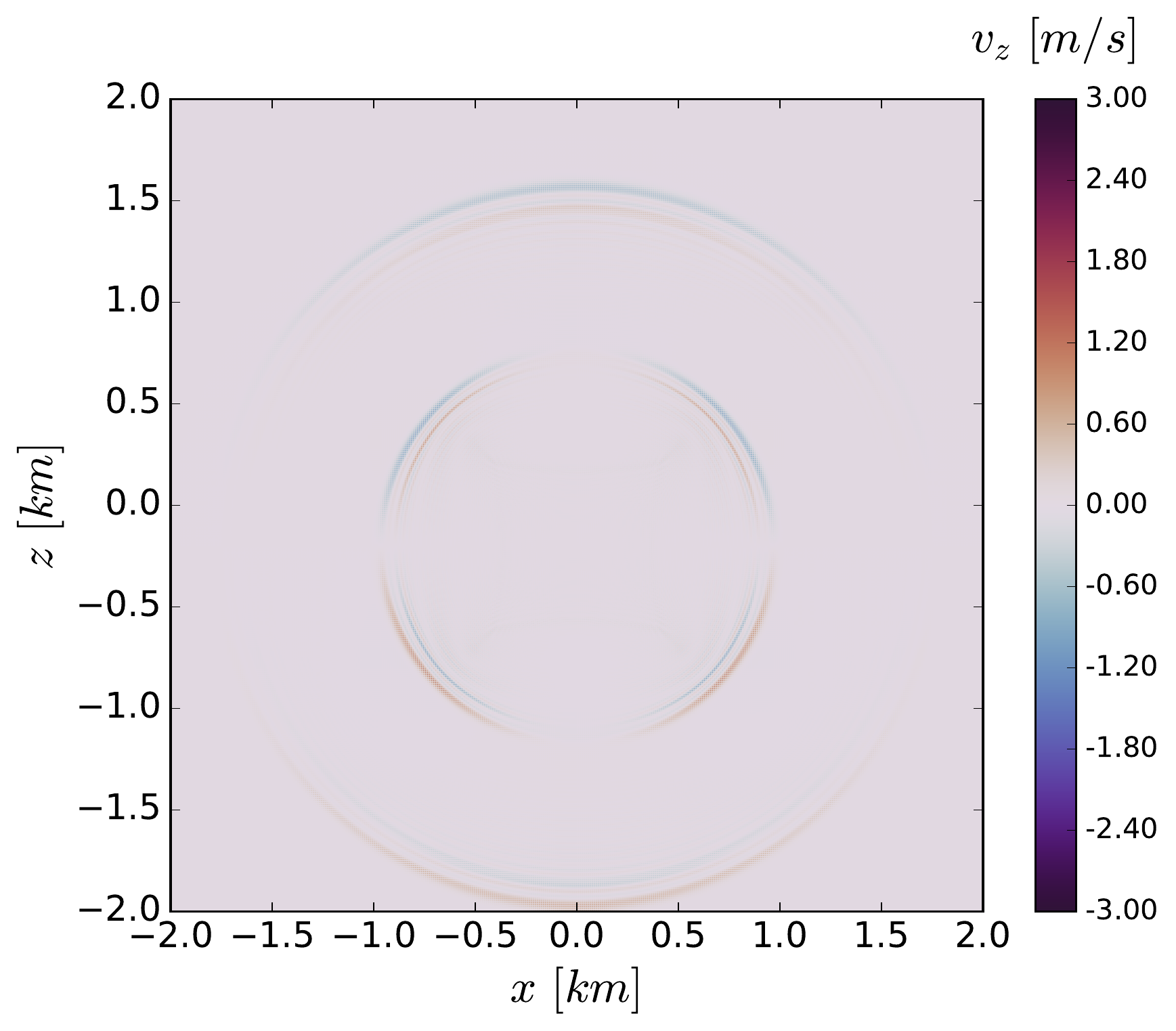}
\caption{Velocity maps for the Ricker wavelet source applied to the normal stresses in homogeneous media. Top-left and top-right figures represent the horizontal and vertical components of the velocity at time $t=0.2$ [s], while bottom-left and bottom-right figures at time $t=0.3$ [s].} 
\label{fig2}
\end{center}
\end{figure}

Fig.\ref{fig1} displays the velocity maps for the source simulated with a sine wavelet in a homogeneous media. Top figures represent horizontal and vertical velocity maps at time $t=0.2$ [s], while bottom figures show their evolution one second later at time $t=0.3$ [s]. On the other hand, Fig.\ref{fig2} shows the velocity maps for the Ricker wavelet source applied to a homogeneous media. In this figure, top panels correspond to horizontal and vertical velocity maps at a time $t=0.2$ [s], while bottom panels at time $t=0.3$ [s]. In the left panels of Fig.\ref{fig1} we can observe two modes, both moving to left and right. The same case can be appreciated in the left panels of Fig.\ref{fig2}. In the right panels of Fig.\ref{fig1} and Fig.\ref{fig2} the modes still appear but the movement is vertical instead of horizontal. The presence of these modes is not surprising, we already knew about their existence thanks to the characteristic analysis realized previously.

Using the same sine wavelet but with different amplitude as the one used for Fig.\ref{fig1}, Cao and Greenhalgh \cite{Cao1992} obtained horizontal and vertical component snapshots for different types of sources: a pure shear energy source, a pure compressional energy source and a combined compressional and shear energy source. As our sources are added to the equations responsible for the evolution of normal stresses ($S_{xx}$ and $S_{zz}$), which are associated with compression, our results are for a pure compressional energy source. Indeed, those results agree with the ones obtained by Cao and Greenhalgh for that kind of source (see Figure 2 from \cite{Cao1992}).

\begin{figure}
\begin{center}
\includegraphics[scale=0.42]{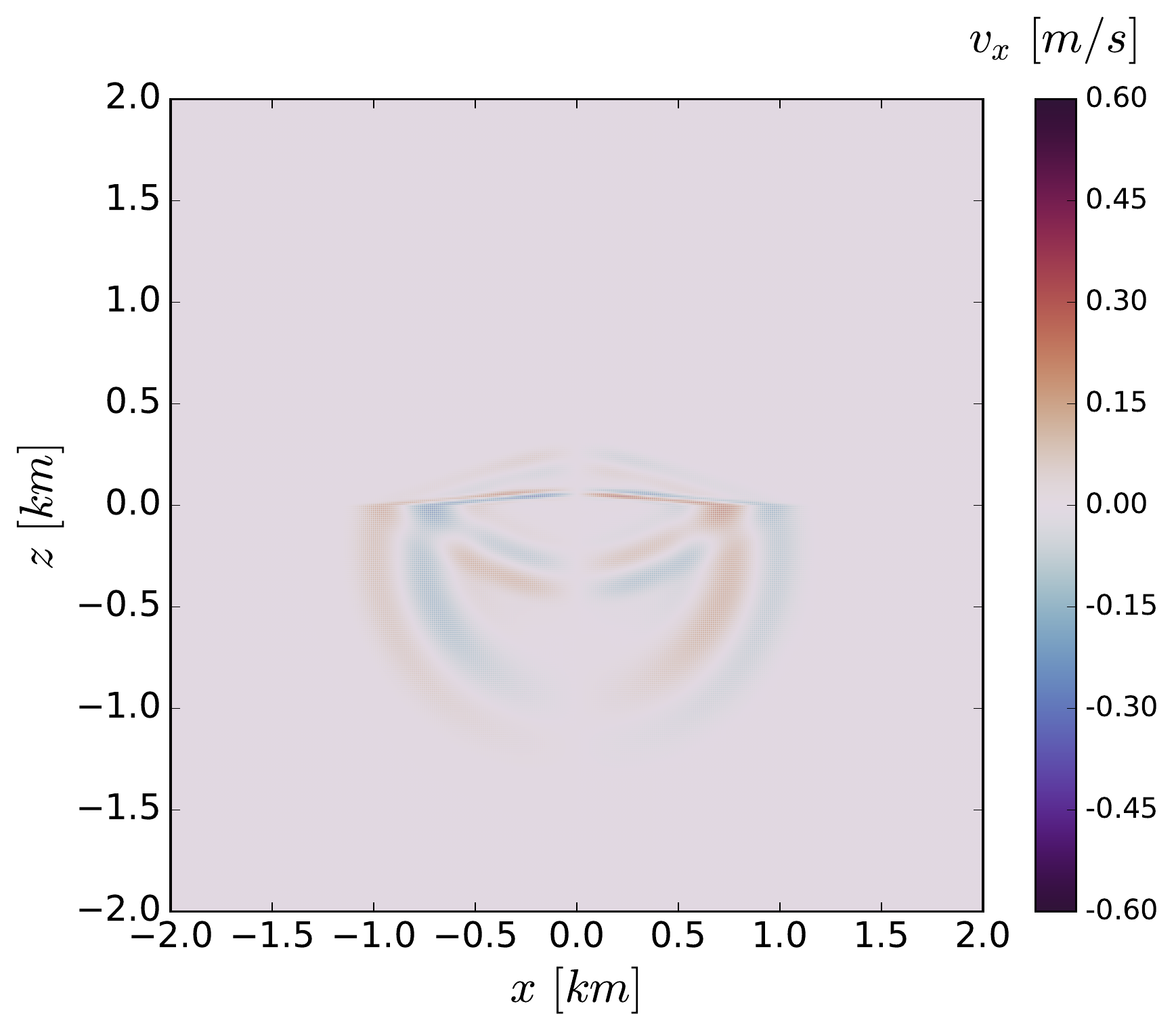}
\includegraphics[scale=0.42]{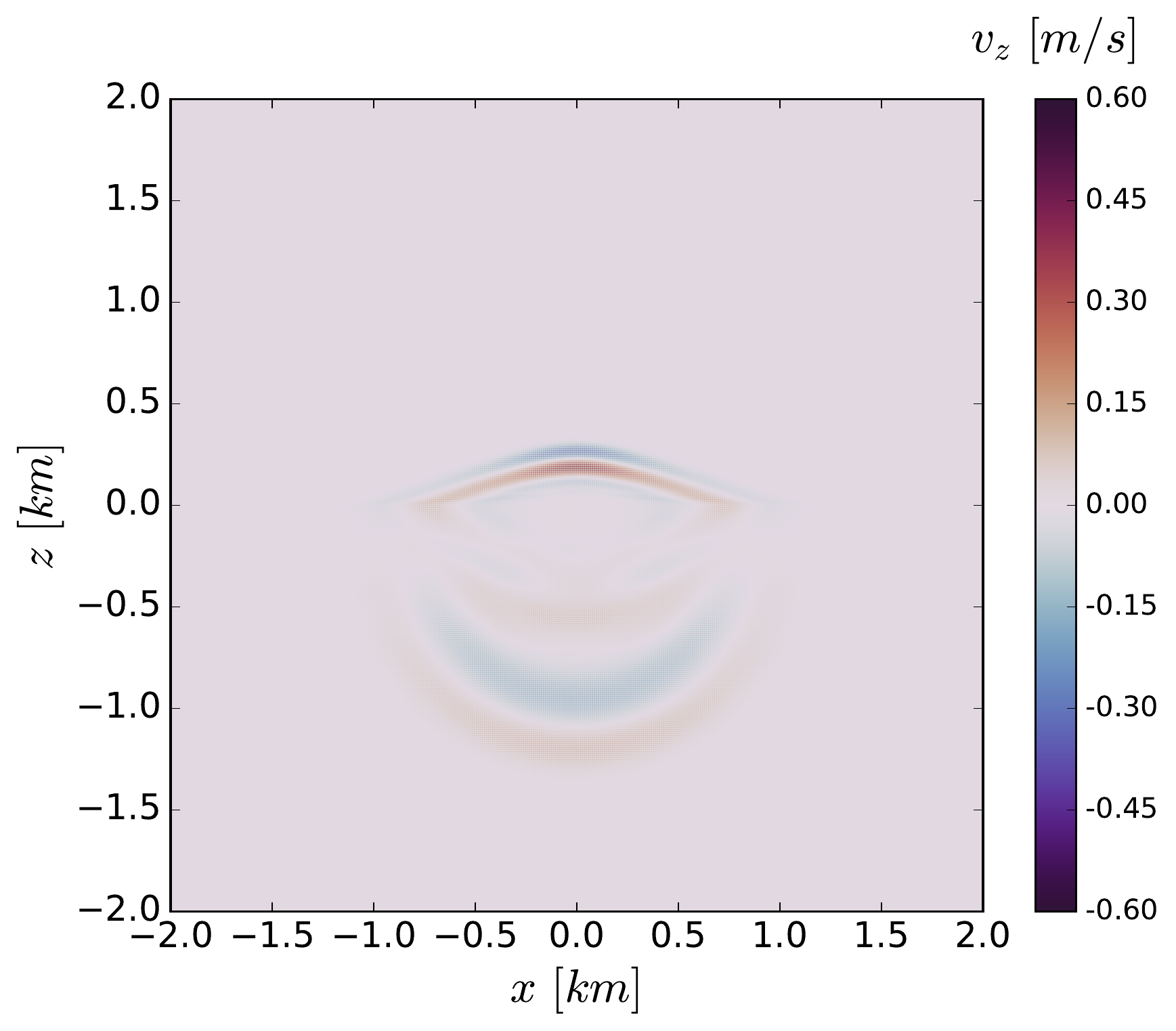}
\includegraphics[scale=0.42]{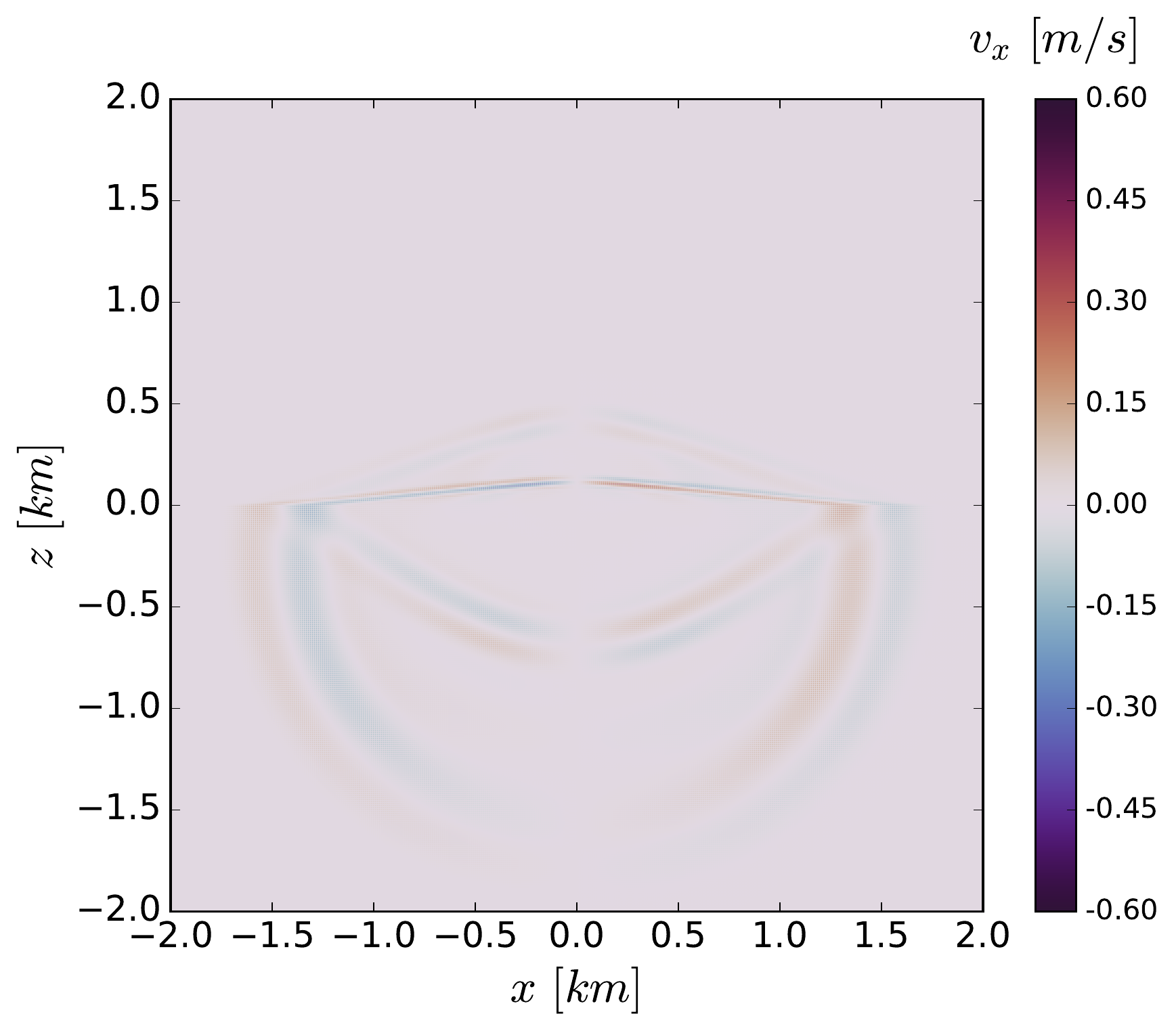} \includegraphics[scale=0.42]{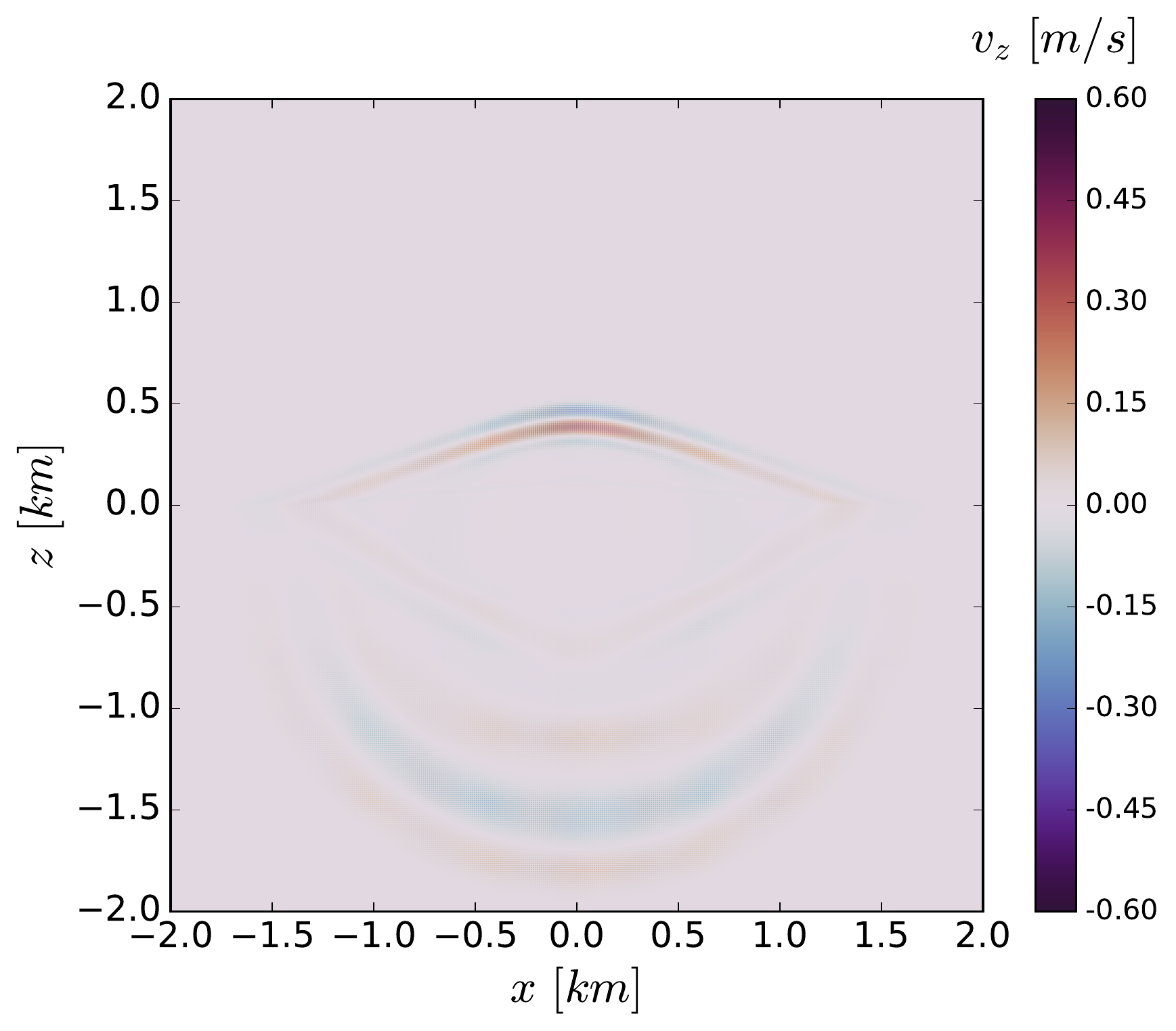}
\caption{Velocity maps for the sine wavelet source applied to the normal stresses in layered media. Top-left and top-right figures represent the horizontal and vertical components of the velocity at time $t=0.2$ [s], while bottom-left and bottom-right figures at time $t=0.3$ [s].}
\label{fig3}
\end{center}
\end{figure}

Fig.\ref{fig3} shows the velocity maps for the source simulated with the sine wavelet in heterogeneous media. The simulation shows the behaviour at the interface between the media, where P- and SV- waves are incident on the internal interface without any kind of boundary conditions. As a consequence, waves are reflected and transmitted to the upper layer where the density is smaller and the velocity wave propagation reduces significantly. For this reason, the media in the upper layer acts as a dissipative material.

\begin{figure}
\begin{center}
\includegraphics[scale=0.42]{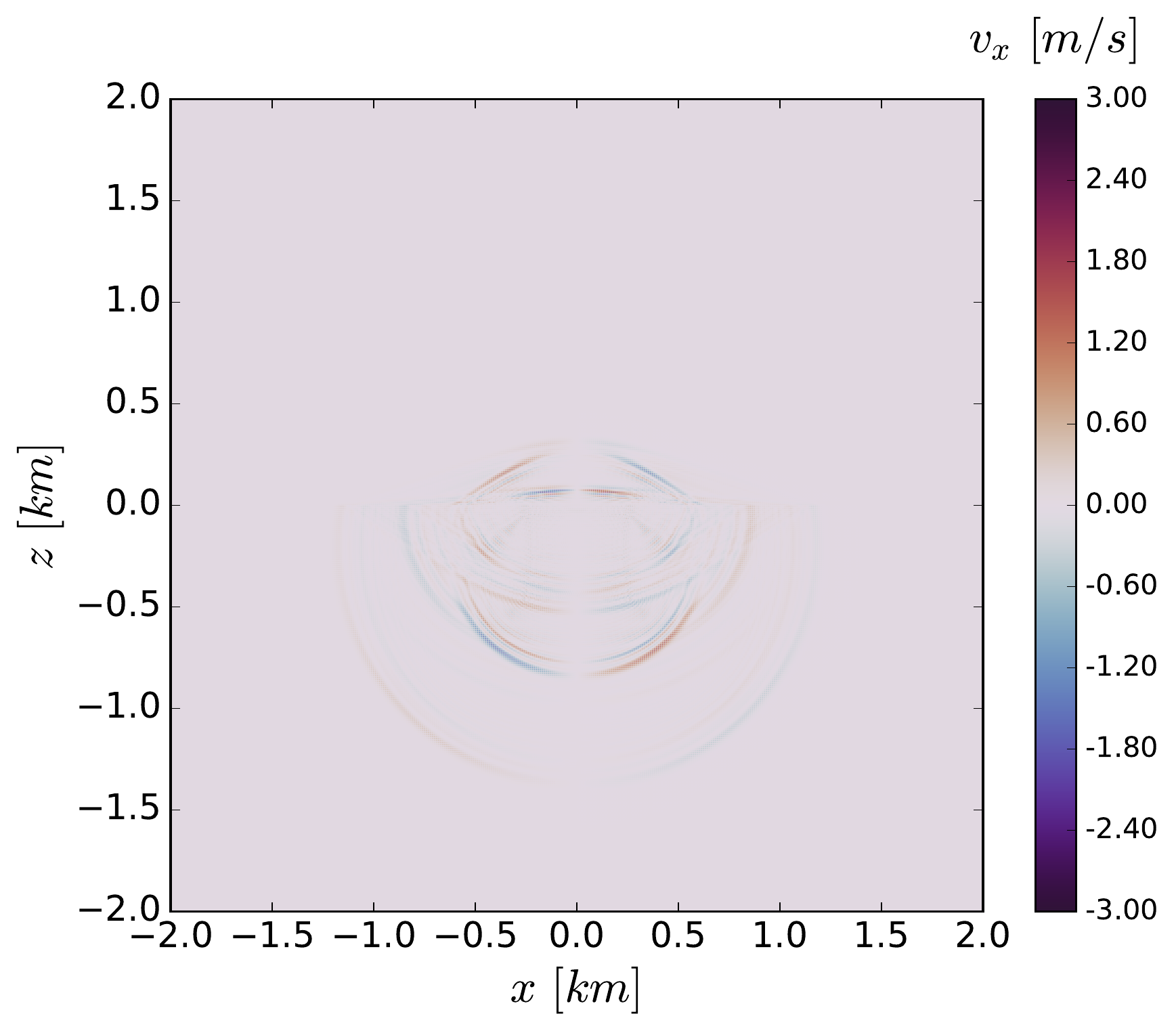}
\includegraphics[scale=0.42]{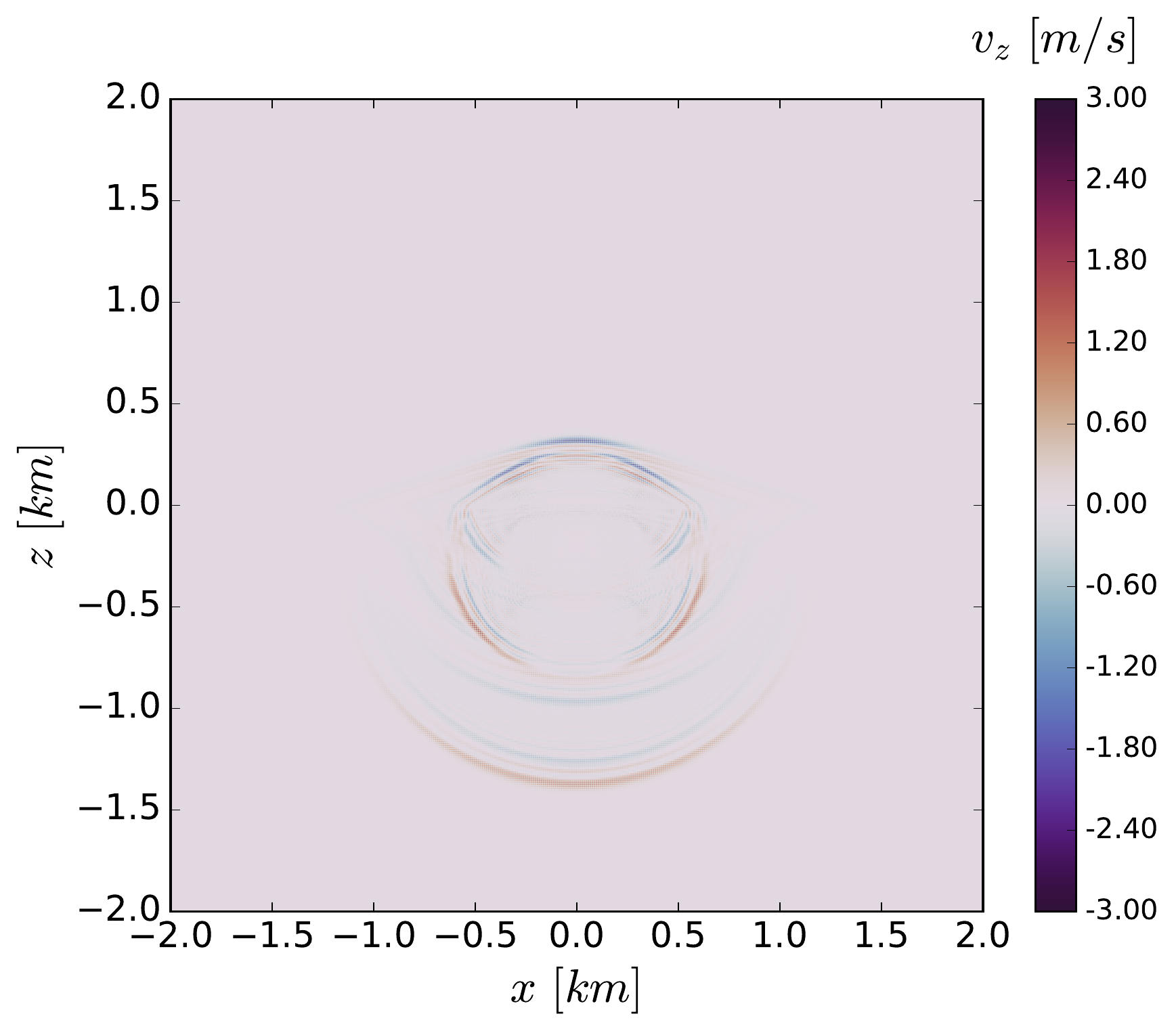}
\includegraphics[scale=0.42]{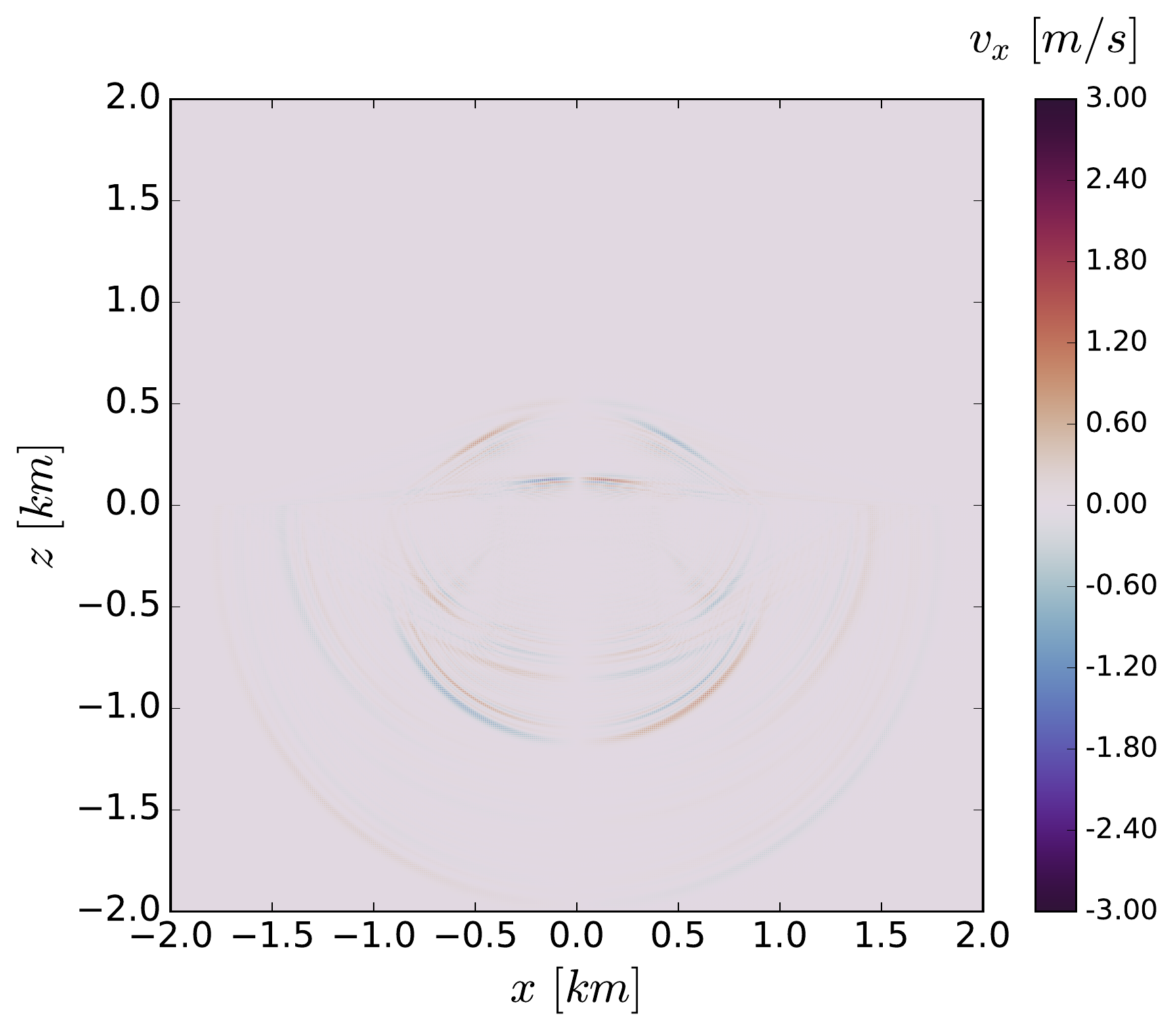}
\includegraphics[scale=0.42]{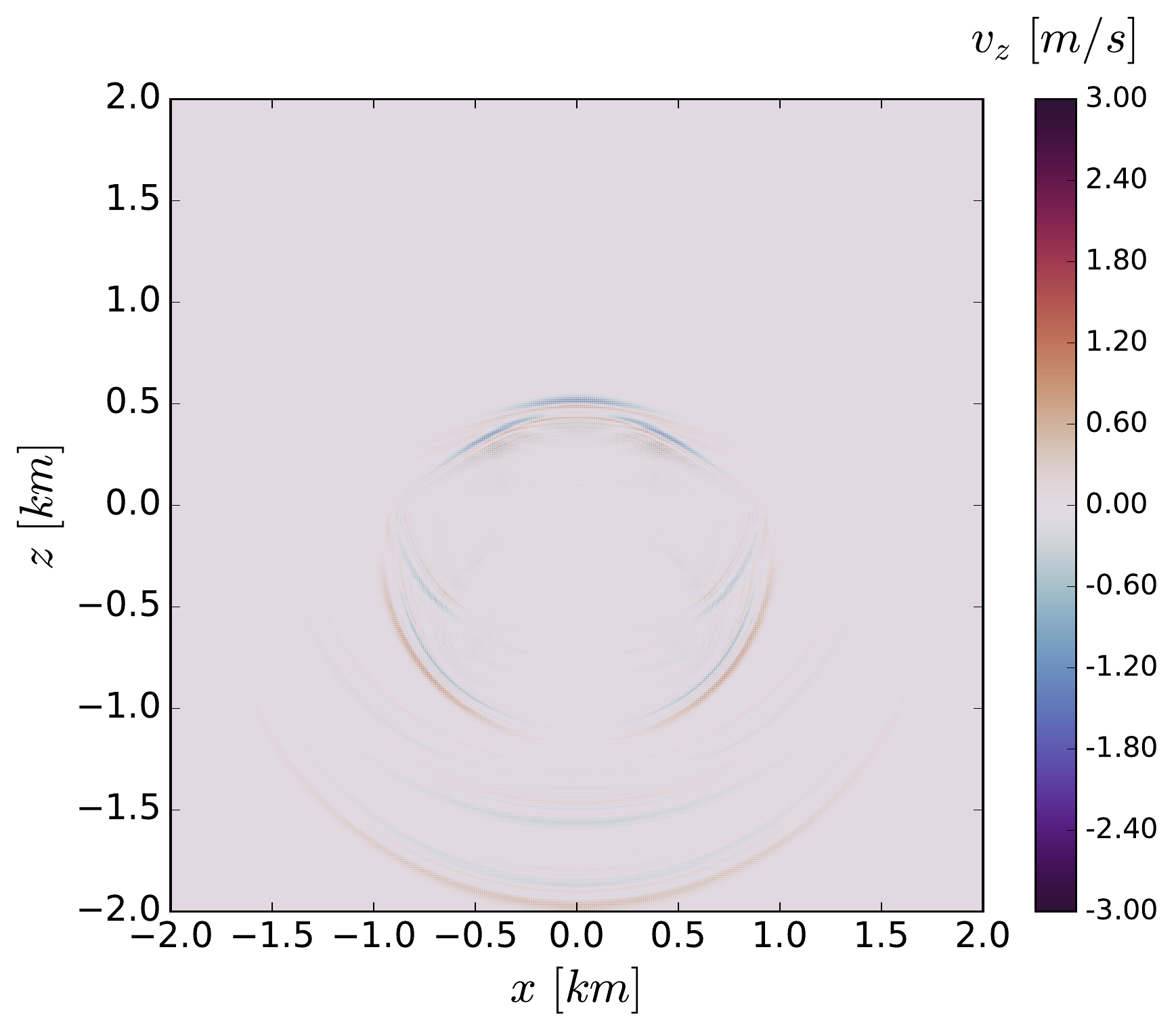}
\caption{Velocity maps for the Ricker wavelet source applied to the normal stresses in layered media. Top-left and top-right figures represent the horizontal and vertical components of the velocity at time $t=0.2$ [s], while bottom-left and bottom-right figures at time $t=0.3$ [s].}
\label{fig4}
\end{center}
\end{figure}

Fig.\ref{fig4} shows the velocity maps for the Ricker wavelet source in heterogeneous media. Again the simulation shows reflection and transmission of waves once they hit the internal boundary. Despite the change in the source model, the upper-medium continues to act as a dissipative media. However, this type of source allows seeing more clearly the several waves that correspond to combinations of P- and SV- waves that are interfering. Moreover, velocity maps of Fig.\ref{fig4} exhibit similar patterns as the ones presented by De Basabe and Sen \cite{deBasabe2015} in their first numerical experiment (see Figure 4 from \cite{deBasabe2015}). However, their snapshots are of the x- and z-components of displacement, not velocity.

\begin{figure}
\begin{center}
\includegraphics[scale=0.3]{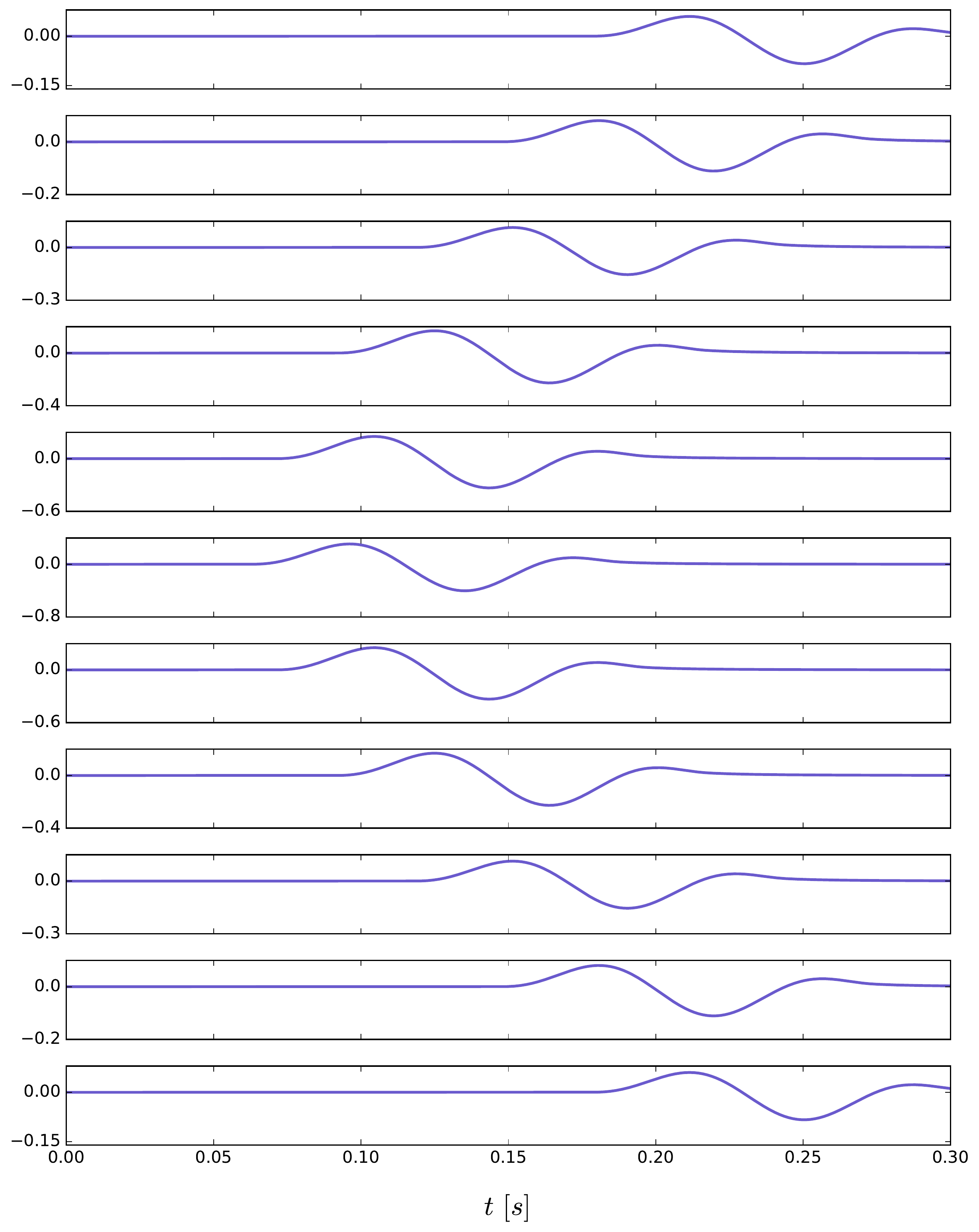}
\includegraphics[scale=0.3]{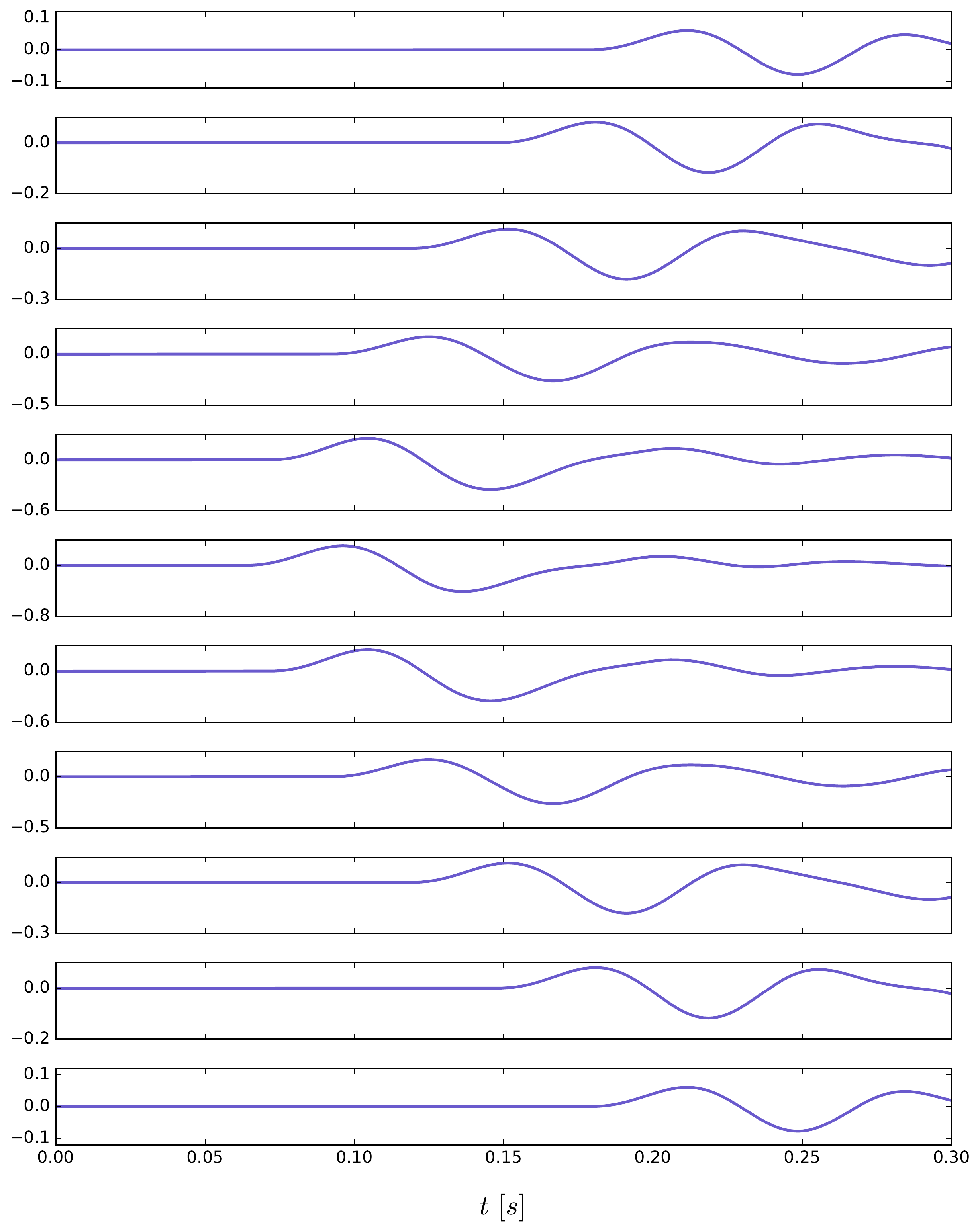}
\caption{Z-velocity time response in homogeneous (left) and layered media (right) for the sine wavelet source. Each plot represents the z-velocity recorded by one of the eleven detectors located between $x=-1.0$ and $x=1.0$.}
\label{fig5}
\end{center}
\end{figure}

\begin{figure}
\begin{center}
\includegraphics[scale=0.3]{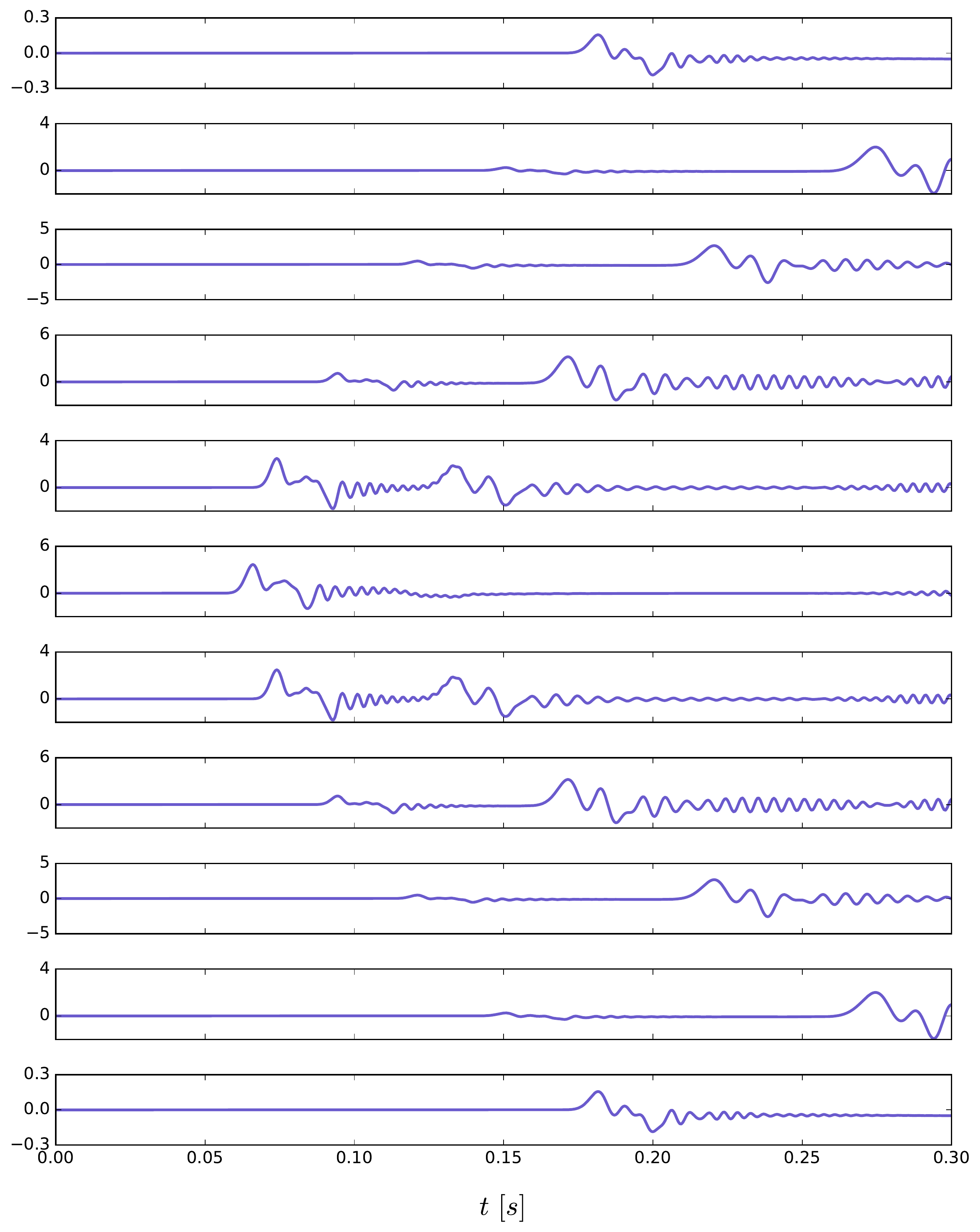}
\includegraphics[scale=0.3]{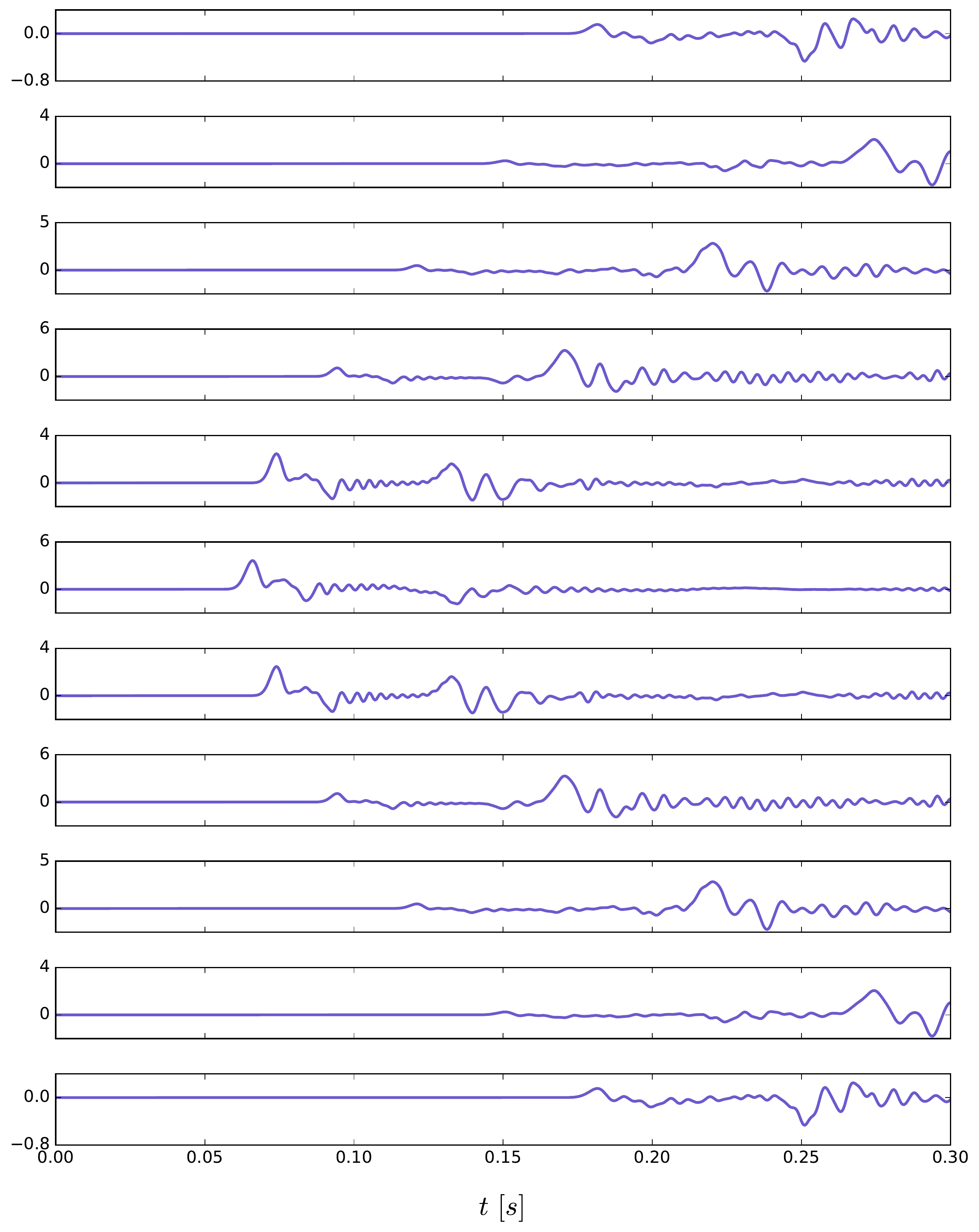}
\caption{Z-velocity time response in homogeneous (left) and layered media (right) for the Ricker wavelet source. Each plot represents the z-velocity recorded by one of the eleven detectors located between $x=-1.0$ and $x=1.0$.}
\label{fig6}
\end{center}
\end{figure}

To compare the behaviour of the sources in homogeneous and heterogeneous media, we plotted the z-velocity as a function of time measured in a specific sequence of points in the domain that act as detectors. More specifically, the z-velocity was recorded by eleven detectors located between positions $x=-1.0$ and $x=1.0$. Fig.\ref{fig5} and Fig.\ref{fig6} show the sine and the Ricker wavelet, respectively, for the homogeneous (left figures) and heterogeneous (right figures) media. Both plots reveal the presence of reflected waves in the inner boundary. However, its presence is more noticeable in the case of the Ricker wavelet (see right panel of Fig.\ref{fig6}) because the reflected waves are recognized in the plots of all detectors, whereas for the sine wavelet (see right panel of Fig.\ref{fig5}) the waves are found particularly around the centre of the domain.

The goal of numerical methods is to find solutions to differential equations at all times.  However, as we show before, when these partial differential equations are solved using finite-differences, it is necessary to discretize them. Then the equations for which we are computing the solution are not exactly those we assume valid in the continuum. As we can see, a truncation error is introduced when we assume the Taylor expansion to a given order. Even though the numerical error associated with the discretized version of a partial differential equation should decrease when the resolution is increased, it is necessary to establish how such error must decrease. Convergence is a notion that relates the decreasing error in terms of the accuracy of the discretization. A difference scheme approximating a partial equation is a convergent scheme at time $t$, if $\lVert \mathbf{u}^{n+1} - \mathbf{v}^{n+1} \rVert \rightarrow 0,$
when $\Delta x \rightarrow 0$ and $\Delta t \rightarrow 0$. Here $\mathbf{u}$ denotes the solution vector of the partial differential equation and $\mathbf{v}$ the solution vector of the PDE evaluated at the grid points \cite{Thomas2013}. As it can be seen, analytically verifying whether a certain approximation can be convergent or not can be quite complex. However, in the numerical case, it is simpler to check that the approximation of the solution converges.

In a  general case, when the exact solution is unknown in the continuum, it is possible to do a Cauchy convergence test computing a given function $f$ with three different resolutions. For instance, consider the resolutions $\Delta x_1 = \Delta x$, $\Delta x_2 = \Delta x/2$ and $\Delta x_3 = \Delta x/4$ and the respective functions computed with each one of these resolutions  $f_1$, $f_2$ and $f_3$. Then we have that the discretization for the second-order finite-difference scheme,  for each function respectively, is given by \cite{Guzman2010} 
\begin{align}
f_1(x) &= f_0(x) + E(x)(\Delta x^2) + \mathcal{O}(\Delta x^3), \\ 
f_2(x) &= f_0(x) + E(x)(\frac{\Delta x^2}{4}) + \mathcal{O}(\Delta x^3), \\ 
f_3(x) &= f_0(x) + E(x)(\frac{\Delta x^2}{16}) + \mathcal{O}(\Delta x^3),
\end{align}
where $f_0$ denotes the exact solution in the continuum and $E$ the error coefficient. From these relations, it is possible to write the following expression, which does not depend on the exact solution
\begin{equation}
    \frac{f_{1}-f_{2}}{f_{2}-f_{3}}=\frac{\frac{3}{4}E(x) \Delta x^{2}+\mathcal{O}(\Delta x^{3})}{\frac{3}{16}E(x) \Delta x^{2}+\mathcal{O}(\Delta x^{3})} \approx 2^2+\mathcal{O}(\Delta x),
\end{equation} 
being the super-index 2 the order of convergence, which in this case corresponds to a second-order (consistent with the order of the finite-difference scheme used).

\begin{figure}
\begin{center}
\includegraphics[scale=0.55]{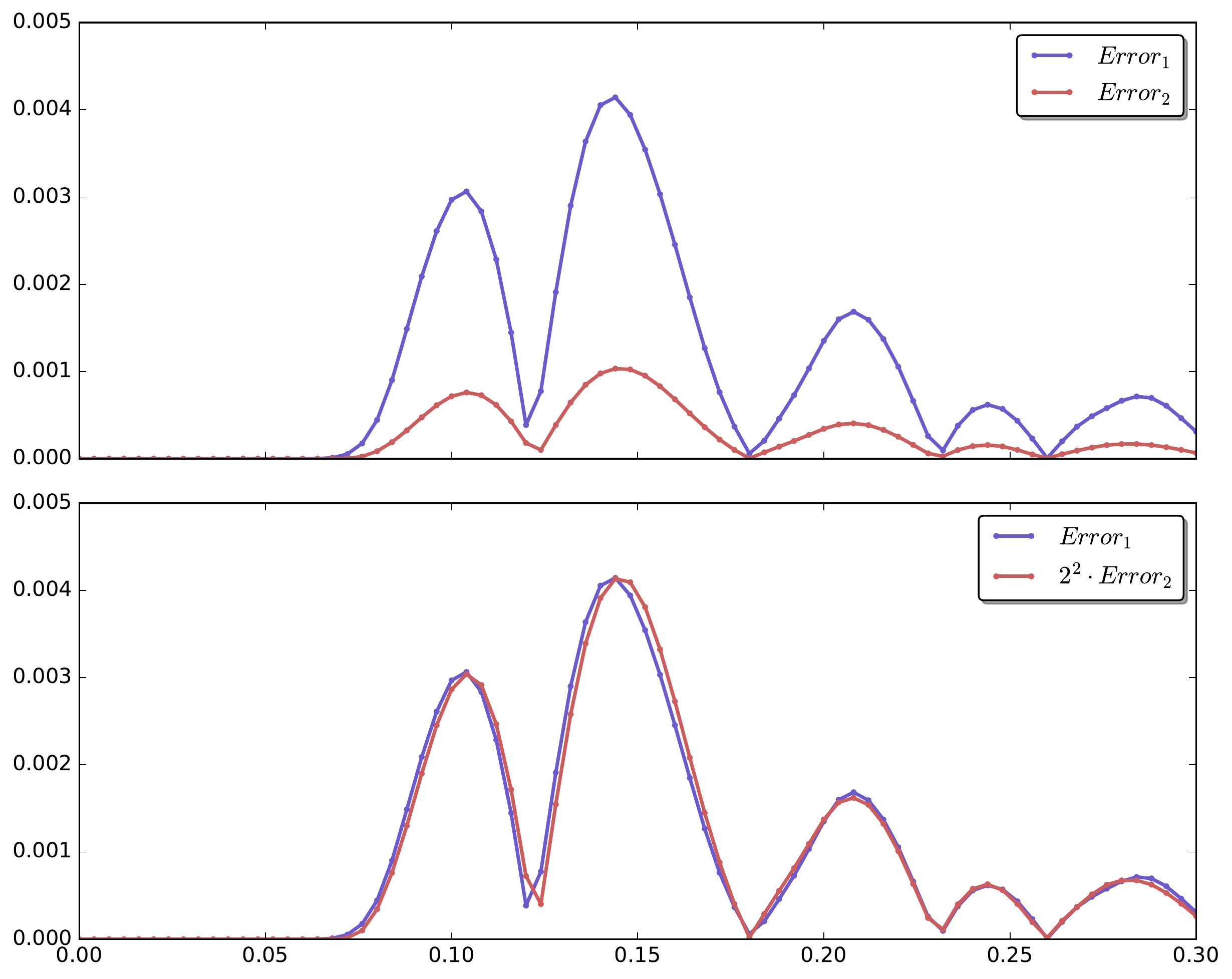}
\caption{Convergence test for the z-velocity measured by one of the detectors in heterogeneous media.}
\label{fig7}
\end{center}
\end{figure}

In Fig.\ref{fig7} we present the convergence test for the z-velocity function, $\dot{u}_{z}$, measured by one of the detectors in heterogeneous media. The resolutions used for the computation of the function are: $\Delta x_1 = \Delta z_1 = 0.01$, $\Delta x_2 = \Delta z_2 = 0.005$ and $\Delta x_3 = \Delta z_3 = 0.0025$, which correspond to $(N_x \times N_z) = (400 \times 400)$, $(N_x \times N_z) = (800 \times 800)$ and $(N_x \times N_z) = (1600 \times 1600)$ numerical points respectively in the simulation box $(-2,-2) \le (x,z) \le (2,2)$. Upper panel of Fig.\ref{fig7} shows the errors $Error_1 = \lvert u_z^1 - u_z^2 \rvert$ and $Error_2 = \lvert u_z^2 - u_z^3 \rvert$, which are computed with the z-velocity function associated to each resolution. As it can be seen the numerical error decreases when the resolution increases, which means that the numerical results are consistent. Additionally, we can obtain the $Error_1$ by multiplying $Error_2$ by the factor $2^2$ (see lower panel), meaning that our numerical simulations converges to second-order, which is consistent with the second-order finite-difference scheme used. It is worth mentioning that independently of the detector the solution always converges.

\section{Discussion} \label{sec:con}

So far we introduced some of the mathematical background required to derive the PDE system for two-dimensional P-SV wave propagation using the velocity-stress formulation. However, our work is mostly focused on the numerical method employed to solve it. We highlight the fact that the literature on the subject is not presented as a straightforward enough recipe that can be followed by undergraduate students taking their first steps in the topic of wave propagation and numerical simulations. For that reason, in this academic paper, we presented step by step how to construct a numerical solution of the system of equations in homogeneous and heterogeneous media using the monolithic approach and a finite-difference scheme.

In the numerical experiments that we performed we compared the velocity maps for two different time-depending sources in homogeneous and heterogeneous media. Our results showed the presence of vertical and horizontal modes that propagate independently of the source but can be more appreciated for homogeneous media using a sine wavelet as the seismic source. From the heterogeneous media simulations, we were able to observe reflection and transmission of waves in the internal boundary. Moreover, we found that the parameters imposed on the upper domain make it acts as a dissipative media.

Similar work was carried out by Cao and Greenhalgh \cite{Cao1992}, who also used a finite difference formulation for the simulation of P-SV-wave propagation in heterogeneous, isotropic media. However, they separated the wave equations into two sets: one for the displacement fields and the other for potential fields. These two sets of equations involve first-order spatial derivatives and require only four equations,  which is one less than Virieux's \cite{Virieux1986} velocity-stress approach used by us and others like Vossen, Robertsson and Chapman \cite{Vossen2002}, and De Basabe and Sen \cite{deBasabe2015}.

Using the velocity-stress approach, recent works were developed by Serdyukov, Koulakov and Yablokov \cite{Serdyukov2019} and Wang, Peng, Lu and Cui \cite{Wang2020}. In the first one, they modelled seismic waves from recorded earthquakes. For 2D simulations they considered a water layer lying on top of solid rock. In water, where S-waves cannot propagate, they defined a constant P-wave velocity. For the solid layer, they defined a constant gradient for the P-wave velocity that increases linearly. There, the S-wave velocity was derived from the last by setting a constant $v_{p}/v_{s}$ ratio. On the other hand, Wang, Peng, Lu and Cui applied the velocity-stress finite-difference method to seismic wave propagation in a fractured medium. With the results they constructed synthetic seismograms, with reflected and transmitted waves that carry fracture information, showing that the fracture can be detected by processing seismic records. 

Hong and Kennett \cite{Hong2003} also modelled seismic waves in heterogeneous media, but using a wavelet-based method. For the validation test, they set a similar domain scheme as the one we used, that is, a two-layered medium with greater elastic wave velocities in the lower layer than those in the upper one, a Ricker wavelet source, an internal boundary and four external boundaries treated with absorbing conditions. The snapshots of elastic wave propagation in the two-layered medium presented in their validation test clearly show how incident waves are reflected and transmitted on the internal boundary. Those snapshots show very similar behaviour to the results we presented in the previous section for the Ricker wavelet in heterogeneous media. However, besides comparing with other works, it is always necessary to carry out a convergence test to check the accuracy of the numerical methods used. In particular, we have found second-order of convergence, which is consistent with the second-order finite-difference scheme implemented, see Fig.\ref{fig7}. 

A study of 3D wave propagation and lateral heterogeneity should be the next step in order to present a fully detailed model. However, many issues arise when more realistic structures are studied trying to transition from 2D to 3D. For example, when lateral heterogeneities take place, gradients of the Lamé parameters should appear in the equation of motion \cite{Alterman1970}, emphasizing that equation (\ref{2.c2}) only takes relevance for a homogeneous media, like our case, where two homogeneous media separated by an interface is presented. When these spatial variations of the parameters are considered the characteristic structure of the system gets complicated, making the system no longer separable into modes. Additionally, in the transition to 3D, the line source becomes obsolete and need to be changed for a point source. This is usually simulated as a delta-like function in space and time \cite{Igel2002}, and also commonly seen as a point body force derived from a seismic moment tensor density distribution \cite{Komatitsch1998, Komatitsch1999}. All these problems that entail more realistic models would be the main focus in a future work.

We conclude by emphasizing that this paper conveyed some of the subtleties that are not found in the literature but are necessary for students, mainly for those taking their first steps in the propagation of waves in complex media, which is well known of great importance in all the areas of physics. This work is also relevant for those who want to improve their computational skills within the framework of a physical problem such as the propagation of seismic waves. It is convenient to highlight that, beyond these computational skills, the present work dealt with important elements of mathematical physics such as tensors and convergence criteria, which are crucial topics for undergraduate physics students. Moreover, this work presented a novel way to teach physics, focusing in wave propagation in heterogeneous media via transversal tools and challenging problems.

\section*{Acknowledgments}
 F. D. L-C acknowledges support from Vicerrectoría de Investigación y Extensión - Universidad Industrial de Santander, Colombia under Grant No. 2493. 
 
\section*{References}
\providecommand{\noopsort}[1]{}\providecommand{\singleletter}[1]{#1}%
\providecommand{\newblock}{}

\end{document}